\documentclass[fleqn,usenatbib]{mnras}

\usepackage{newtxtext,newtxmath}

\usepackage[T1]{fontenc}
\usepackage{ae,aecompl}

\usepackage{graphicx}	
\usepackage{amsmath}	
\usepackage{amssymb}	

\usepackage{threeparttable}
\usepackage{longtable}
\usepackage{multicol}
\usepackage{multirow}
\usepackage{etaremune}
\usepackage{array}
\usepackage{gensymb}

\title[Probing the extragalactic fast transient sky at minute timescales with DECam]{Probing the extragalactic fast transient sky at minute timescales with DECam}

\author[I. Andreoni et al.]{\noindent I. Andreoni$^{1,2,3,4,\text{*}}$, 
J. Cooke$^{1,3,5}$,
S. Webb$^{1,3}$,
A. Rest$^{6,7}$,  
T. Pritchard$^{1,8}$,
M. Caleb$^{1,9,10}$,\newauthor
S.-W. Chang$^{3,5,9}$, 
W. Farah$^{1}$, 
A. Lien$^{11,12}$,
A. M{\"o}ller$^{5,9}$,
M.~E. Ravasio$^{13,14,1}$ , \newauthor
T.~M.~C. Abbott$^{15}$,
S. Bhandari$^{1,16}$, 
A. Cucchiara$^{17}$, 
C. Flynn$^{1}$,  
F. Jankowski$^{10,1,5}$, \newauthor
E.~F. Keane$^{18}$,
T.~J. Moriya$^{19}$,  
C.~A. Onken$^{5,9}$, 
A. Parthasarathy$^{1,3,16}$,
D. C. Price$^{1,20}$, \newauthor
E. Petroff$^{21,22}$, 
S. Ryder$^{2,23}$,
D. Vohl$^{22}$,
C. Wolf$^{5,9}$\\
$^1$Swinburne University of Technology, PO Box 218, Hawthorn, VIC 3122, Australia\\
$^2$Australian Astronomical Observatory, 105 Delhi Rd, North Ryde, NSW 2113, Australia\\      
$^3$ARC Centre of Excellence for Gravitational Wave Discovery (OzGrav), Australia\\
$^4$Division of Physics, Math and Astronomy, California Institute of Technology, Pasadena, CA 91125, USA\\
$^5$ARC Centre of Excellence for All-sky Astrophysics (CAASTRO)\\
$^6$Space Telescope Science Institute, 3700 San Martin Dr., Baltimore, MD 21218, USA\\
$^{7}$Department of Physics and Astronomy, Johns Hopkins University, 3400 North Charles Street, Baltimore, MD 21218, USA\\
$^8$Center for Cosmology and Particle Physics, New York University, 726 Broadway, New York, NY 10004, USA\\
$^{9}$Research School of Astronomy and Astrophysics, Australian National University, Canberra, ACT 2611, Australia\\
$^{10}$Jodrell Bank Centre for Astrophysics, School of Physics and Astronomy, The University of Manchester, Manchester M13 9PL, UK\\
$^{11}$Center for Research and Exploration in Space Science and Technology (CRESST) and NASA Goddard Space Flight Center, \\Greenbelt, MD 20771, USA\\
$^{12}$Department of Physics, University of Maryland, Baltimore County, 1000 Hilltop Circle, Baltimore, MD 21250, USA\\
$^{13}$Universit\`a degli Studi di Milano-Bicocca, Dipartimento di Fisica U2, Piazza della Scienza 3, Milano 20126, Italy\\
$^{14}$INAF - Osservatorio Astronomico di Brera, via Bianchi 46, Merate 23807, Italy\\
$^{15}$Cerro Tololo Inter-American Observatory, National Optical Astronomy Observatory, Casilla 603, 1700000 La Serena, Chile\\
$^{16}$CSIRO Astronomy and Space Science, Australia Telescope National Facility, PO Box 76, Epping NSW 1710, Australia\\
$^{17}$College of Science and Mathematics, University of the Virgin Islands, \#2
Brewers Bay Road, Charlotte Amalie, USVI 00802\\
$^{18}$SKA Organisation, Jodrell Bank Observatory, Macclesfield SK11 9DL, UK\\
$^{19}$Division of Theoretical Astronomy, National Astronomical Observatory of Japan, National Institutes of Natural Sciences,\\
2-21-1 Osawa, Mitaka, Tokyo 181-8588, Japan\\
$^{20}$Department of Astronomy,  University of California Berkeley, Berkeley CA 94720\\
$^{21}$Anton Pannekoek Institute for Astronomy, University of Amsterdam, Amsterdam, the Netherlands\\
$^{22}$ASTRON, Netherlands Institute for Radio Astronomy, Dwingeloo, the Netherlands\\
$^{23}$Department of Physics \& Astronomy, Macquarie University, NSW 2109, Australia,\\
$^{\text{*}}$E-mail: andreoni@caltech.edu
}

\date{Submitted for publication in MNRAS}

\pubyear{2019}

\begin{document}
\label{firstpage}
\pagerange{\pageref{firstpage}--\pageref{lastpage}}
\maketitle

\begin{abstract}
Searches for optical transients are usually performed with a cadence of days to weeks, optimised for supernova discovery.  The optical fast transient sky is still largely unexplored, with only a few surveys to date having placed meaningful constraints on the detection of extragalactic transients evolving at sub-hour timescales.  Here, we present the results of deep searches for dim, minute-timescale extragalactic fast transients using the Dark Energy Camera, a core facility of our all-wavelength and all-messenger Deeper, Wider, Faster programme.  We used continuous 20\,s exposures to systematically probe timescales down to 1.17\,minutes at magnitude limits $g > 23$ (AB), detecting hundreds of transient and variable sources. Nine candidates passed our strict criteria on duration and non-stellarity, all of which could be classified as flare stars based on deep multi-band imaging. Searches for fast radio burst and gamma-ray counterparts during simultaneous multi-facility observations yielded no counterparts 
to the optical transients. Also, no long-term variability was detected with pre-imaging and follow-up observations using the SkyMapper optical telescope. We place upper limits for minute-timescale fast optical transient rates for a range of depths and timescales.  Finally, we demonstrate that optical $g$-band light curve behaviour alone cannot discriminate between confirmed extragalactic fast transients such as prompt GRB flashes and Galactic stellar flares.   
\end{abstract}



\section{Introduction}

The optical transient sky is largely unexplored at short timescales. Most successful time-domain surveys aim at detecting supernovae and variable events evolving on week or month timescales. Those include, for example, the Supernova Legacy Survey \citep{Astier2006}, the Cal\'an/Tololo Survey \citep{Hamuy1999}, the Palomar Transient Factory \citep[PTF,][]{Rau2009}, the Catalina Sky Survey \citep{Drake2009}, Pan-STARRS \citep{Stubbs2010}, the Dark Energy Survey \citep[DES,][]{DES2016}, the All-Sky Automated Survey for Supernovae \citep[ASAS-SN,][]{Shappee2014, Holoien2017}, and now the Asteroid Terrestrial-impact Last Alert System (ATLAS\footnote{\url{http://atlas.fallingstar.com/}}) and the Zwicky Transient Facility \citep[ZTF\footnote{\url{https://www.ztf.caltech.edu/}},][]{Bellm2019ZTF,Graham2019ZTF} among others.

Recent work unveiled new classes of luminous, rapidly-evolving supernovae \citep{Poznanski2010Sci, Kasliwal2010, Drout2014, Shivvers2016, Arcavi2016, Rodney2018, Pursiainen2018, De2018Sci} performing observations with nightly and sub-nightly cadence.  \cite{Rest2018} present the most extreme of these luminous fast transients discovered to date, which shows a rise time of 2.2~days, a time above half-maximum of only 6.8~days, and a peak luminosity comparable to Type\,Ia supernovae. Bright optical flashes have also been observed during rapid follow up of long gamma-ray bursts using robotic telescopes \citep{Fox2003,Cucchiara2011,Vestrand2014,Martin-Carrillo2014, Troja2017opt}.

Rapid optical and infrared transients of high interest are kilonovae, which are associated with gravitational-wave events \citep[e.g.,][]{Abbott2017MMA} in addition to short gamma-ray bursts
\citep{Perley2009,Tanvir2013a,Berger2013k,Gao2015,Jin2015,Jin2016,Troja2018KN,Jin2019NatAs}.  
The discovery of a kilonova counterpart to the neutron-star merger GW170817 allowed the precise pin-pointing of the event in the sky
\citep{Coulter2017, Soares-Santos2017, Valenti2017, Arcavi2017GW, Lipunov2017,Tanvir2017} and allowed more than 70 facilities to monitor its evolution at many wavelengths \citep[e.g.,][]{Abbott2017MMA}. Now we know that multiple components characterise the emission arising from mergers such as GW170817 \citep[see for example][]{Cowperthwaite2017,Kasliwal2017,Villar2017}. In particular, observations of this transient revealed an early, blue component that evolves in about three days (with a rising phase of hours), but its origin is still unclear. Early detection and monitoring of a population of kilonovae can allow us to understand the nature of this rapidly evolving component \citep{Arcavi2018}. 
As multi-messenger astronomy grows in importance, more and more surveys are dedicated to the search for kilonova-like transients \citep[e.g.,][]{Doctor2017} or to probe the background of contaminant sources during the follow up of gravitational-wave triggers \citep{Cowperthwaite2018}.  

In addition to dedicated observing campaigns, new searches for fast optical transients are performed in archival data of surveys such as DES \citep{Pursiainen2018arXiv} and PTF \citep{Ho2018}.  In the latter, the authors discovered an already-identified gamma-ray burst afterglow that was identified independently from gamma-ray triggers \citep{Cenko2015}. 
The Sky2Night project \citep{vanRoestel2019} searched for fast transients using PTF, with observations performed with 2\,hr cadence for 8 nights.  \cite{vanRoestel2019} place upper limits on rates of 4\,hr and 1\,day timescale transients at $R<$ 19.7 limiting magnitude, obtaining $R < 37 \times 10^{-4}$deg$^{-2}$d$^{-1}$ and $R < 9.3 \times 10^{-4}$deg$^{-2}$d$^{-1}$, respectively. 

Only a few wide-field surveys have been carried out at timescales shorter than 1\,hr. The continuous 30-minute cadence of the {\it Kepler} K2 project led to the first discovery of the optical shock breakout of a core-collapse supernova \citep[][but see also \citet{Rubin2017}]{Garnavich2016}. If considered independently from the long-lasting supernova emission, this constitutes a rare example of hour-timescale extragalactic fast transient detection. \cite{Bersten2018} present the remarkable discovery of another optical supernova shock breakout, revealing an increase by $\Delta$M$_V \sim 0.6$ in $\sim$~25~minutes and estimated to last $\sim 0.1$\,day.  Other fast-cadenced surveys include a monitoring of the Fornax galaxy cluster \citep{Rau2008}, and blind surveys such as ROTSE III \citep{Rykoff2005}, the Deep Lens Survey \citep[DLS,][]{Becker2004}, MASTER \citep{Lipunov2007}, and Pi of the Sky \citep{Sokolowski2010}. Results from the Pan-STARRS Medium-Deep Survey were reported by \cite{Berger2013}. The authors provided a summary of the upper limits on extragalactic fast optical transients evolving on $\sim$0.5 hours timescales until 2013: the upper limits placed by all those surveys (Fig.\,\ref{fig:surveys}) confirm that Galactic M-dwarf flares outnumber extragalactic fast transients by a large factor \citep[up to several orders of magnitude,][]{Rau2008}. More recent surveys explore similar short-timescale regimes, for example the High Cadence Transient Survey \citep[HiTS]{Forster2016}. Interesting detections of fast-rising transients, increasing their luminosity by $>1$\,magnitude in the restframe near-ultraviolet wavelengths between two consecutive nights \citep{Tanaka2016}, fuels the field of fast transient searches with exciting prospects.

Current surveys such as ZTF and Catalina Sky Survey can provide data suitable to search for minute to hour timescale transients.  In the near future, the Large Synoptic Survey Telescope \citep[LSST,][]{LSST2009}
is expected to come online.  LSST will survey the sky to deep magnitude limits and thanks to its large field of view ($\sim10$\,deg$^2$) it is expected to discover several thousands of extragalactic transients every night.  The choice of the observing cadence will determine the degree to which LSST can contribute to different research areas in time domain astronomy.    

This work aims to explore a new region of the optical luminosity-timescale phase space \citep[e.g.,][]{Cenko2017}, focusing on the search for faint extragalactic transients fully evolving in minutes.  We performed deep and fast-cadenced observations with the Dark Energy Camera \citep[DECam,][]{Flaugher2015}, a wide-field imager mounted at the prime
focus of the 4-m Blanco telescope at the Cerro Tololo Inter-American Observatory (CTIO) in Chile. Such observations were performed in the framework of the Deeper Wider Faster programme\footnote{\url{http://dwfprogram.altervista.org/}} \citep[Cooke et al., in prep;][]{Andreoni2018DWF}, described in the next section. Fast cadenced observations with DECam are a key optical component of DWF that enables new studies, including the search for counterparts to fast radio bursts \citep[FRBs,][]{Lorimer2007a} at many other wavelengths. FRBs are transients detected at radio wavelengths as dispersed signals that last only a few milliseconds. Several arguments support an extragalactic origin for FRBs, including the identification of the host galaxy of the repeating burst FRB\,121102 \citep{Tendulkar2017}.  However, the nature of FRBs is still unknown, thus the detection of possible optical or high-energy counterparts could shed light on the FRB physics.

A subset of the total quantity of optical data collected during DWF campaigns is analysed and presented in this paper (Table\,\ref{tab:target fields}-\ref{tab:log}). The chosen observations allow the most systematic analysis of minute-timescale fast transients over multiple nights, as the observing conditions were the most uniform of the full dataset.

The paper is organised as follows. Observations are presented in Section\,\ref{sec: observations}.  Our analysis is described in Section\,\ref{sec: search for optical bursts} and its results are presented in Section\,\ref{sec: results}. Searches for longer duration optical transients, FRBs, and GRBs are presented in Section\,\ref{sec: multi-wavelength}. Rates for fast optical transients in the survey-depth transient-timescale phase space are presented in Section\,\ref{sec: rates}.  We discuss the results in Section\,\ref{sec: discussion} and we conclude with a summary of this work in Section\,\ref{sec: summary}. 

\section{Observations}
\label{sec: observations}

We describe the Deeper Wider Faster programme in Sec.\,\ref{subsec: DWF}, the criteria driving the choice of the target fields in Sec.\,\ref{subsec: target fields}, and the characteristics of the unique, fast-cadenced optical images in Sec.\,\ref{subsec: fast-cadenced images}.

\subsection{The Deeper Wider Faster programme}
\label{subsec: DWF}

The DWF programme is designed to unveil the fastest and most elusive bursts in the sky. Identifying counterparts to FRBs constitute the primary goal of the programme. The novelty of the approach adopted by the DWF team resides in coordinating deep, wide-field, fast-cadenced observations {\it simultaneously} with multiple small to large all-messenger facilities. By contrast,
most of the existing efforts to accomplish the same science goals rely on different observing strategies.  Usually when an interesting transient is detected by the survey telescope (or neutrino and gravitational-wave detectors), then a network of facilities receives the trigger and reacts to follow up the transient. Such a {\it reactive} approach has several limitations, for example the multi-wavelength information is usually collected hours or days after the first detection.  This causes the loss of possibly significant information or, in the case of FRBs, may be the cause of the lack of any counterpart discovered to date\footnote{\cite{Hardy2017} report upper limits on optical flux at times coincident with bursts from the repeating FRB~121102. The extremely fast cadence of Thai~National~Telescope/ULTRASPEC (70.7~ms frames) makes their results meaningful, however we caution that {\it i)} possible optical emission could have been fainter than the 5$\sigma$ upper limit of ULTRASPEC ($m_i\sim 15.75$, AB), and {\it ii)} FRB~121102 may not be representative of the whole FRB population. }.  During DWF campaigns the fields are observed at multiple wavelengths in a {\it proactive} way: before, during, and after minute- and sub-minute-timescale fast transients shine.  Rapid detection and prompt and long-term follow up of transients is key to the success of the programme. Blanco/DECam has been the core optical facility during the first five DWF observing runs (2 pilot and 3 operational runs), spanning between January 2015 and February 2017. 

By 2018, the DWF programme has grown and now has more than 40 participating facilities, including the Subaru/Hyper~Suprime-Cam (HSC) used as the core optical instrument for simultaneous observations in February 2018.  The thorough analysis of the Subaru and the multi-wavelength data will be presented in future publications.

\subsection{Target fields}
\label{subsec: target fields}

Coordinating multiple telescopes that shadow each other constrains the region of sky observable during simultaneous observations. Such limitations become particularly significant when the observatories are located on different continents.  In fact, the ground-based core facilities used during DWF from 2015 to 2017 for simultaneous or rapid follow-up observations are located in Chile (Blanco/DECam, Rapid Eye Mount telescope, Gemini-South) and in Australia (Parkes, Molonglo, ATCA, and SkyMapper).  
Constraints change when using core facilities other than DECam and Australian radio telescopes for DWF observations, for example when using Subaru/HSC in Hawaii, or the MeerKAT radio telescope in South Africa in the near future. For space-based observatories such as {\it Swift}, constraints include limited time on fields far from the poles, earth occultation times, and Sun constraints. 

We demonstrated that these specific geographical constraints can be overcome (Cooke et al., in prep) and successful DWF observations from Chile and Australia can be performed all year around. We chose the target fields in relation to the following criteria:  
 
\begin{itemize}

\item Sky locations where FRBs were previously discovered (here FRB131104). The field of view of the 13-beam receiver at Parkes (see Section\,\ref{sec: Parkes and Molonglo}) well matches the DECam field of view (FoV) but, in addition, the localisation error for those FRBs discovered with Parkes ($\sim$15$'$ diameter) allows targeted, simultaneous observations using telescopes with FoV smaller than DECam, such as REM and {\it Swift}/UVOT-XRT;
\item Nearby galaxy clusters (e.g., Antlia), nearby galaxies, or globular clusters;

\item Legacy fields (e.g., COSMOS field) having dense photometric and spectroscopic information in multiple wavelengths and space-based high resolution imaging; and/or fields where we have previous DECam deep imaging and colour information, located at high Galactic latitude and observable for $\gtrsim$1\,hr from Chile and Australia (Prime, 3hr, 4hr).

\end{itemize}

\begin{table}
	\centering
	\caption[Equatorial and Galactic coordinates of the target fields.]{Equatorial (J2000) and Galactic coordinates of the target fields selected for this work, from a larger DWF dataset. 
    }
	\label{tab:target fields}
	\begin{tabular}{lcccc} 
		\hline
		Target Field & RA & Dec &{\it l} & {\it b}  \\
		\hline
		 3hr   & 03:00:00 & $-$55:25:00 & 272.47784$\degree$ &	$-$53.43243$\degree$ \\
		 4hr  &04:10:00 & $-$55:00:00 & 264.94437$\degree$ & $-$44.75641$\degree$   \\
		 Prime  & 05:55:07 & $-$61:21:00 & 	270.35527$\degree$ & $-$30.26267$\degree$  \\
		 FRB131104  & 06:44:00 & $-$51:16:00 & 260.53726$\degree$ & $-$21.94809$\degree$  \\
		 Antlia & 10:30:00 & $-$35:20:00  & 272.94307$\degree$ & $-$19.17249$\degree$ \\
		\hline
	\end{tabular}
\end{table}

More than 15 fields were observed with high-cadence simultaneous observations during DWF observing runs. Table\,\ref{tab:target fields} reports the coordinates of target fields whose data are analysed in this work. 

\subsection{Fast-cadence imaging with DECam}
\label{subsec: fast-cadenced images}

DWF observing campaigns have produced a large quantity of multi-wavelength data. These are analysed in real time or near-real time to search for FRBs, their possible counterparts, and to discover Galactic and extragalactic fast transients. DWF collected $\gtrsim 10,000$ optical images with DECam, mainly in the $g$ filter, which allows $\sim 0.5$\,mag deeper observations in comparison with other filters. The expected depth for 20\,s exposure in $g$-band is 23.7\,magnitudes (AB), against 22.6, 23.1, 22.6, and 21.6 magnitudes of the $u$-$r$-$i$-$z$ filters, respectively (1.0 arcsec FWHM seeing).   Typical seeing and relatively high airmass ($\sim$1.5) required for the coordinated observations can affect these values, sometimes moving the $g$-band limiting magnitude closer to $\sim$23 mag. 

The observing strategy with DECam during DWF runs is based on a series of continuous exposures. Our experience has shown that a 20\,s exposure time is optimal to 1) enable sub-minute timescale variability exploration, 2) reach individual image depth $\geq 23$ to observe large sky volumes, 3) enable efficient data transfer and image subtraction in real time, and 4) probe a range of depths and durations when images are analyzed individually and in various stacked forms.  

Telescope movements are limited to a few arcseconds in order to cover the largest common area of the sky with adjacent exposures. 
In this work we analyse 25.76\,hr of high-cadenced images: 20\,s continuous exposures in $g$ band, separated by $\sim$30\,s where CCDs complete the readout and the new exposures start. Each image covers an effective field of view of 2.52~deg$^2$, accounting for CCD gaps and the frame area cropped during the alignment of the images. Details of the observations discussed in this work are presented in Table\,\ref{tab:log}. Data are processed and calibrated with the NOAO High-Performance Pipeline System \citep{Valdes2007, Swaters2007}.

\begin{figure}
\centering
	\includegraphics[width=\columnwidth]{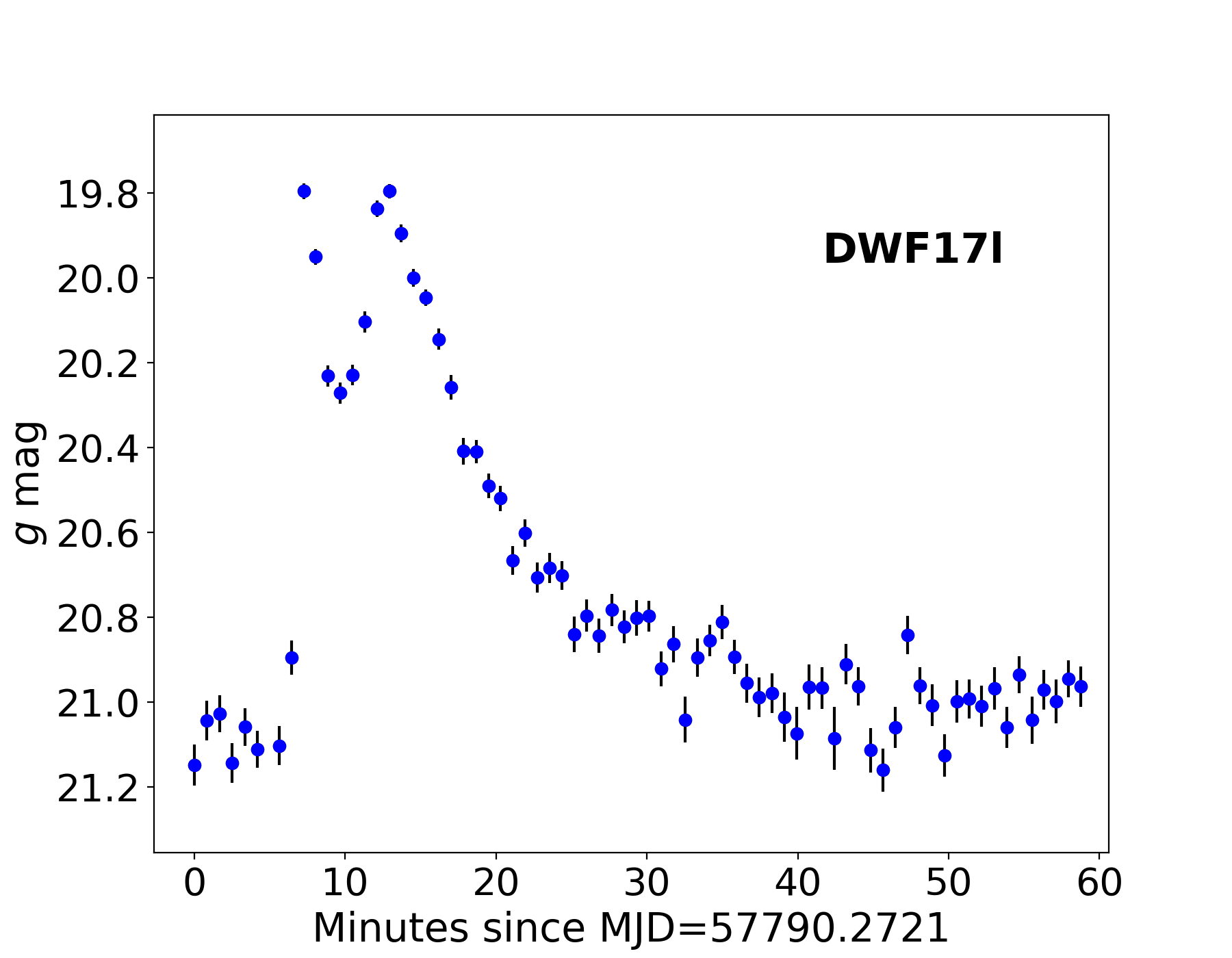}
    \includegraphics[width=\columnwidth]{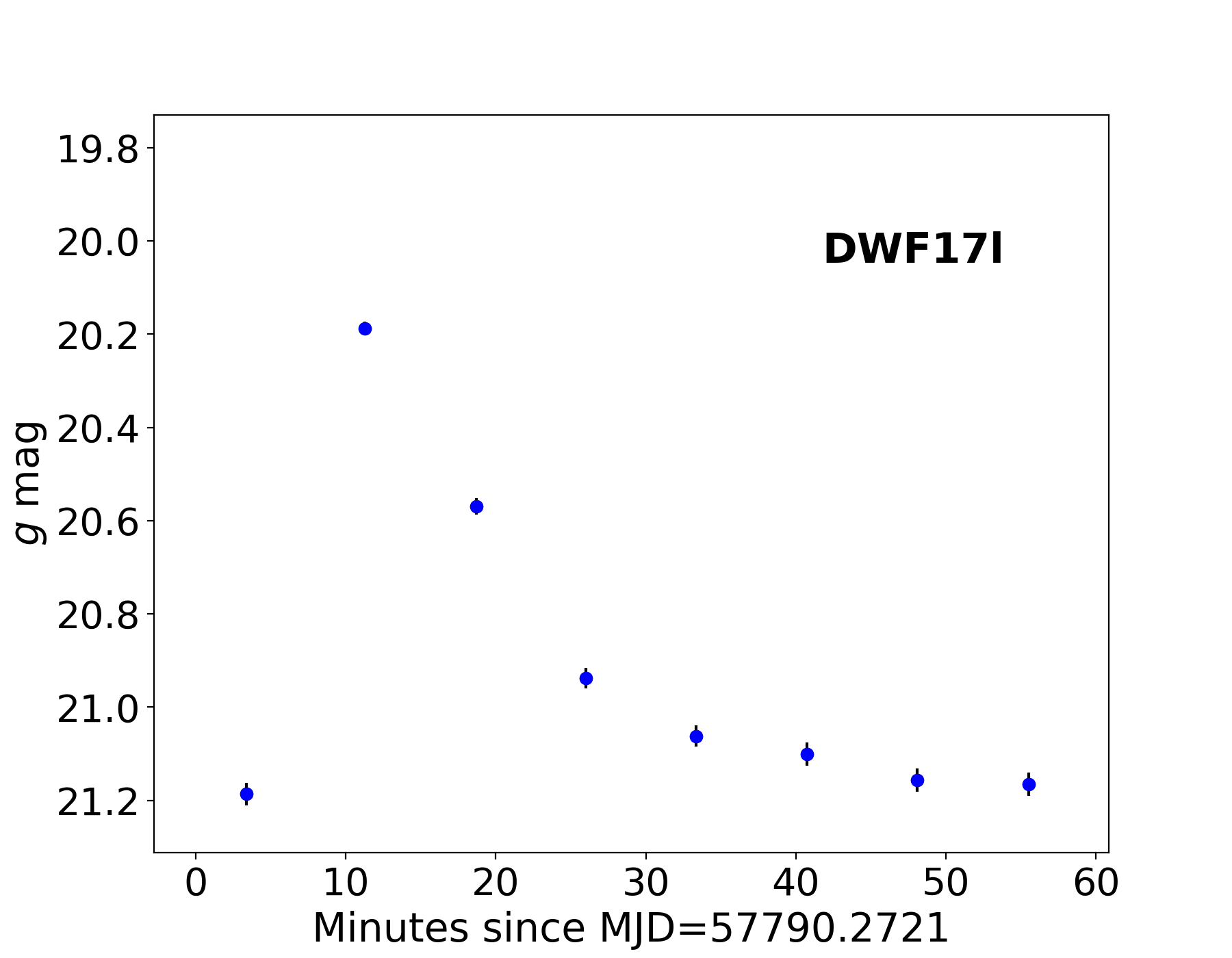}
    \caption[The multi-peak flare DWF17l.]{ The multi-peak structure of DWF17l is visible in light curves built with series of 20\,s exposure images (top panel), but is lost when stacking sets of 9 images (bottom panel).  Even if some time resolution is lost, stacking of images acquired during DWF observations allows deeper and more fast-cadenced searches than most existing surveys. The source DWF17l is located at coordinates RA= 6:42:20.333, Dec= --52:10:24.78 (J2000) and the colour of its quiescent counterpart suggests the flare to have arisen from a M5 red-dwarf star. } 
    \label{fig: lc example double peak}
\end{figure}

\begin{figure}
\centering
    \includegraphics[width=\columnwidth]{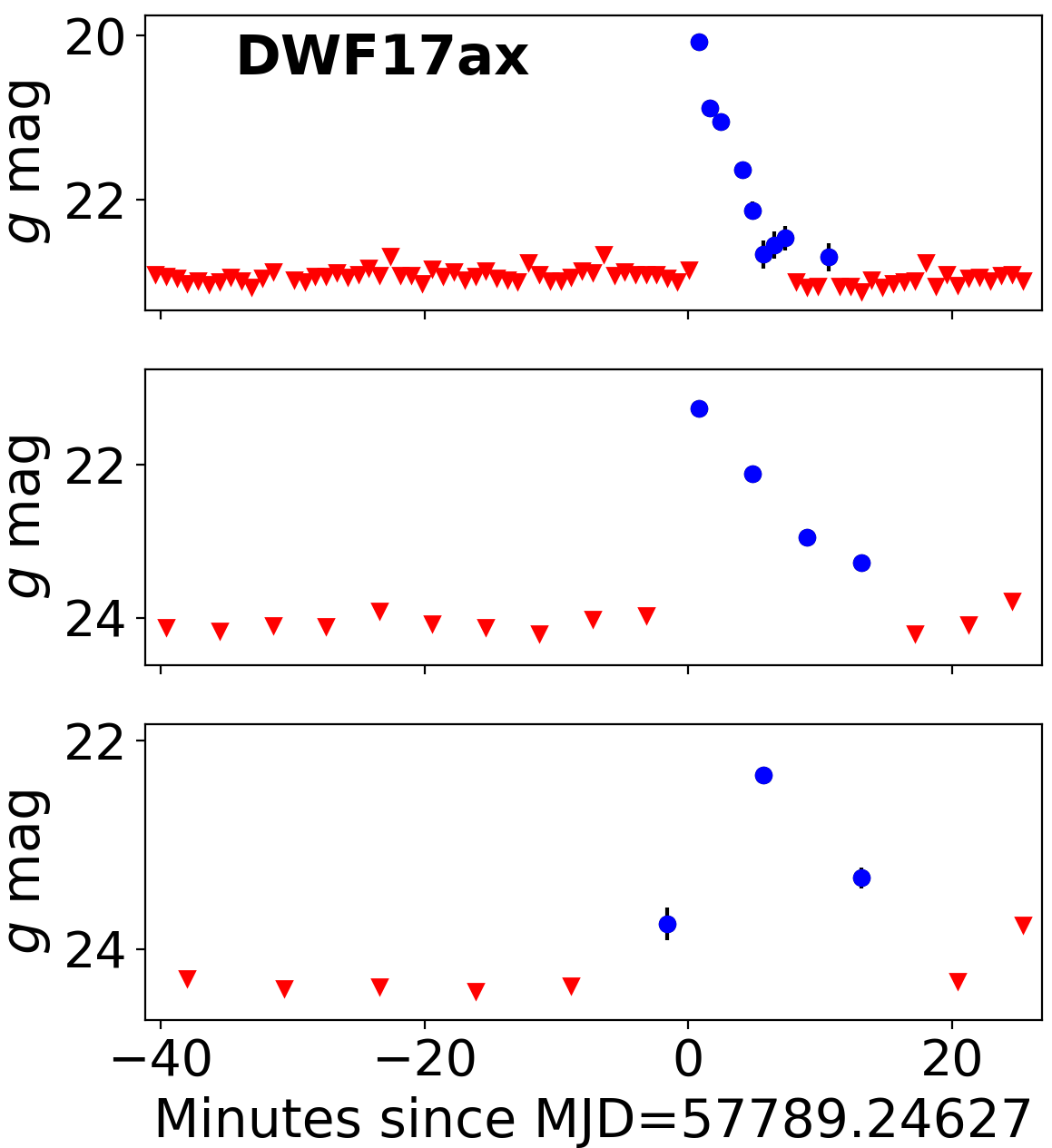}
    \caption[Light curve of the fast-transient candidate DWF17ax]{Light curves of the fast-transient candidate labelled DWF17ax (whose discovery images are shown in Figure~\ref{fig: stamp detection}) obtained with individual 20-s exposures ({\it top}) and by stacking sets of 5  ({\it centre}) and 9  ({\it bottom}) fast-cadence images. Red triangles indicate 5$\sigma$ upper limits. } 
    \label{fig: lc example DWF17ax}
\end{figure}

The high cadence of our images allows us to study transient and variable events in great detail.  For example, the structure of the stellar flare DWF17l in Figure\,\ref{fig: lc example double peak}, sampled at 50\,s intervals (20\,s exposure $+$ 30\,s overhead), is lost when median-stacking sets of 9 consecutive images, equating to $\sim 7$\,min intervals.  Similarly, fast-cadence imaging reveals a non-monotonic fade of the light curve of DWF17ax, difficult to study at slower cadence (Figure\,\ref{fig: lc example DWF17ax}).

\begin{table}
	\centering
	\caption[Time (expressed in hours) for which the target fields were observed.]{Total time (expressed in hours) for which the target fields were observed.}
	\label{tab:log}
	\begin{tabular}{lccccc} 
		\hline \hline
		Date & \multicolumn{5}{c}{Target Field} \\
		YYMMDD & 3hr  & 4hr & Prime & FRB131104 & Antlia \\
		
		\hline
		151218 & 0.98  & 1.18  & - & - &-   \\
		151219 & 1.21  & 1.24  & - & - &-  \\
	    151220 & 1.40  & 0.91 & - & - & -  \\
	    151221 & 1.47 & 1.05  & - & - &-  \\
	    151222 & 1.66 & 0.23  & - & - & -  \\
	    170202 & - & - & 0.97 & 0.96 & 0.86  \\
	    170203 & - & - & 1.35 & 0.58 & 1.00  \\
	    170205 & - & - & 1.11 & 0.87 &  0.72  \\
	    170206 & - & - & 0.96 & 0.98  & 1.00  \\
	    170207 & - & - & 1.03 & 1.02 &  1.02 \\	   
	     
		\hline
	\end{tabular}
\end{table}

\section{Analysis: search for extragalactic fast transient candidates}
\label{sec: search for optical bursts}

Data are searched using the custom {\it Mary} pipeline \citep{Andreoni2017mary}.  The pipeline identifies optical transients with image subtraction techniques, processing all CCDs in parallel with the Green II (g2) supercomputer at Swinburne University of Technology (now superseded by the OzStar supercomputer).

\begin{figure}
\centering
	\includegraphics[width=\columnwidth]{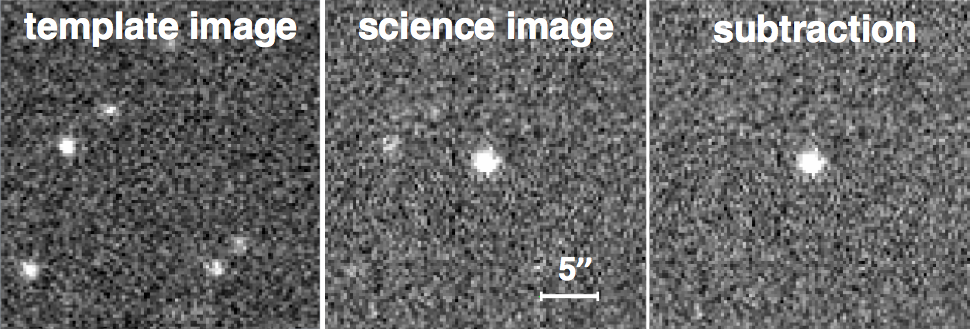}
    \caption[Example of detection `postage-stamp' images.]{Example of detection `postage-stamp' images.  The extragalactic fast transient candidates (eFTC, in this case DWF17ax) is present in the science image ({\it center}) and is not present at the same coordinates in the deeper template images ({\it left}). The image subtraction between science and template images leaves a bright, PSF-shaped residual brighter than the background ({\it right}). The light curve of DWF17ax is shown in Figure~\ref{fig: lc example DWF17ax}. } 
    \label{fig: stamp detection}
\end{figure}

The {\it Mary} pipeline automatically outputs a list of candidates and generates three small `postage stamp' images for each candidate for visual inspection (see for example Figure\,\ref{fig: stamp detection}). In addition, a companion code generates aperture photometry light curves centred at the location of the discoveries, with radius $1.5 \times$\,FWHM measured from nearby stars. Light curves presented in this paper are first calibrated against the all-sky USNO-B1 catalog (which provides a number of Southern-Hemisphere sources large enough to enable calibration of individual CCDs independently) and then calibrated against the AAVSO Photometric All Sky Survey \citep[APASS,][]{Henden2016} catalogue, that provides AB system measurements. 
Our tests indicated the $g$-band magnitudes obtained this way to be consistent with magnitudes from the Sloan Digital Sky Survey catalogue (where there is overlap with DWF fields) within $\lesssim 0.1$ magnitudes.  

Template images used for the analysis are always deeper than the individual 20\,s exposure images, however their limiting magnitude is usually different for different fields, depending on the availability of archival images from previous observations. For the identification of short-timescale optical transients
it is not necessary to have template images older than a few minutes, however the availability of deep templates obtained ad much earlier/later times can greatly help the classification process.
When fast transients are identified, we perform a more accurate classification of the detections, based on the photometry performed on deep images, multi-filter information, and object information found when cross-matching with existing catalogues. Asteroids are easily identifiable thanks to the fast cadence of the observations, which makes the movement of the objects in the sky evident when comparing a chronological sequence of images.

First, we search for transient and variable sources in 1870 individual 20s-exposure images, taken with regular cadence. Second, we stack the images in sets of 5, 9, 13, and 17 images to reach deeper magnitude limits and search for fainter fast-evolving transients.

\begin{figure}
\centering
	\includegraphics[width=\columnwidth]{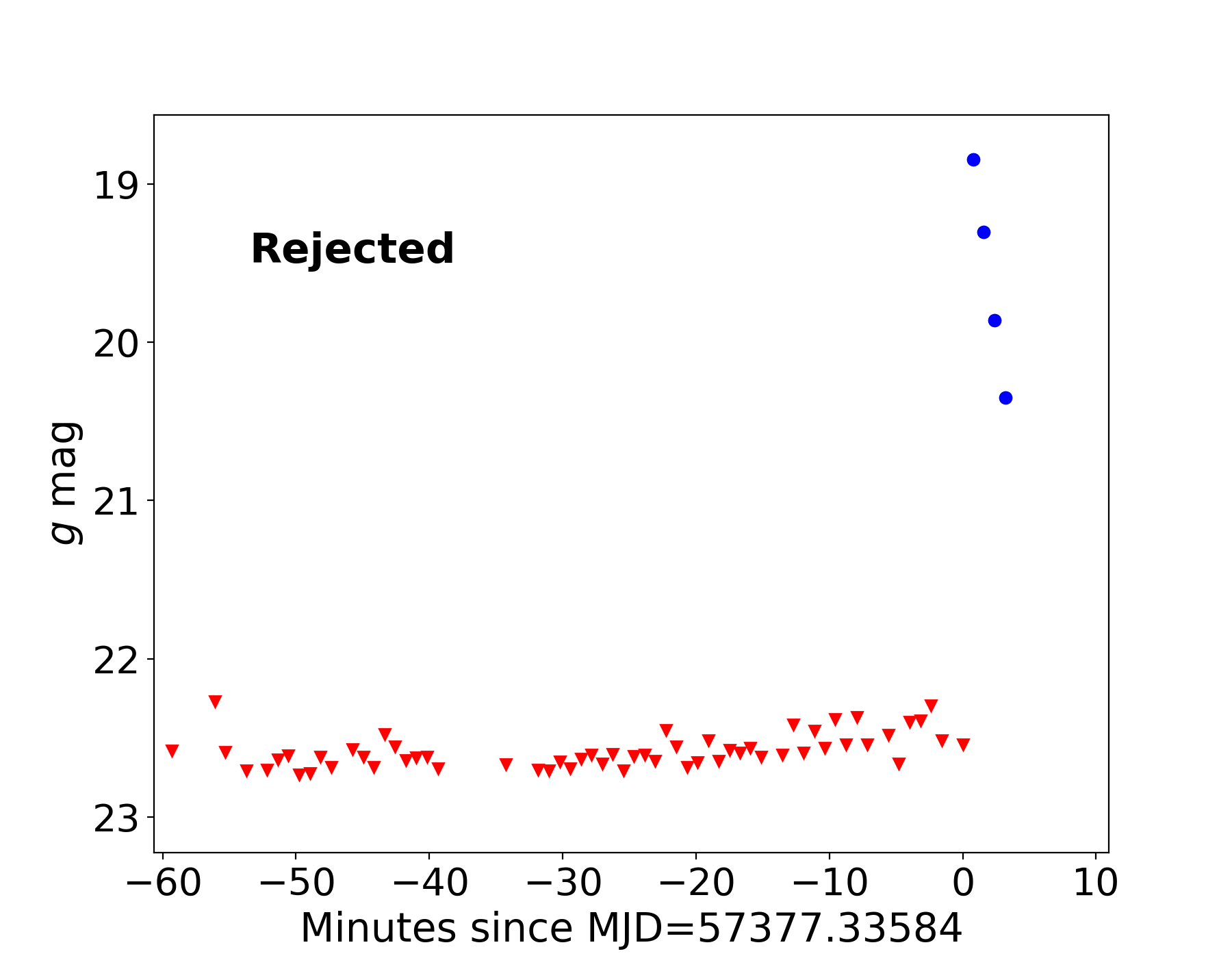}
    \caption[Light curve of a transient that did not fulfill our selection criteria. ]{Light curve of an eFTC that did not fulfill the selection criteria described in Section\,\ref{sec: search for optical bursts}.  In particular, the candidate was detected in the last image acquired on the observing night in which it was discovered, but we require at least one epoch of non-detection at the beginning and at the end of each observing night to better constrain the transient duration.  
    } 
    \label{fig: lc example ignored}
\end{figure}

\subsection{Selection criteria}
\label{subsec: selection criteria}

 \begin{figure*}
 \centering
	\includegraphics[width=2.\columnwidth]{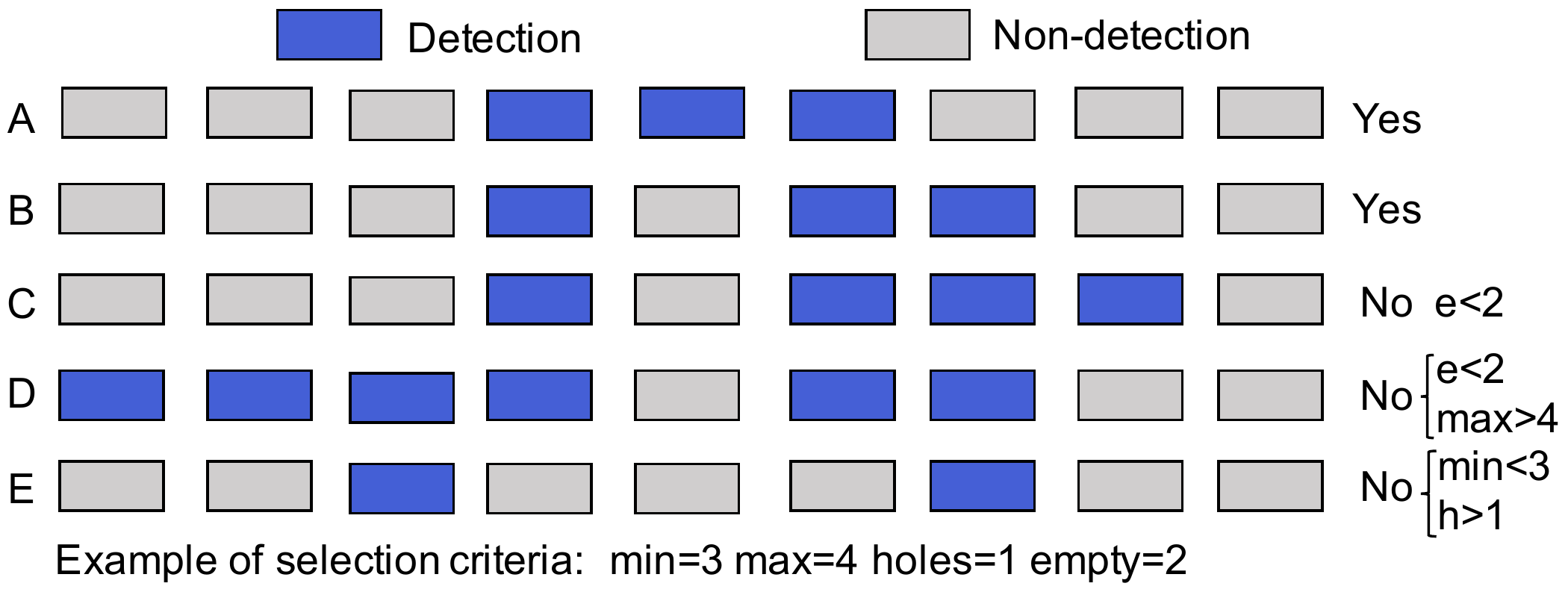}
    \caption[Example of selection criteria]{Example of selection criteria application. Given a set of consecutive images (9 images in this example, but on the order of hundred images per night during DWF observations), we define the following selection criteria: at least 3 detections (min=3), at least 2 images without detection at the beginning and at the end of the observing night of that target field (empty=2), at most 4 detections (max=4), and at most 1 image without detection between the first and last detection (holes=1). We apply up to 9 collections of such selection criteria for each set of images in order to keep a nearly constant ratio of minimum number of detections and holes (see Table\,\ref{tab:minmax detections}). }
    \label{fig:selection criteria}
\end{figure*}

At the end of the processing, candidates automatically identified with the {\it Mary} pipeline are selected aiming at identifying extragalactic fast transient candidates (eFTCs).  In particular, we search for astrophysical transient events that evolve at minutes timescales and are not spatially coincident with Galactic sources. 

When searching for eFTCs we consider only the first night that a detection occurs for those objects detected in more than one night. Upon first detection of a luminosity increase (using image subtraction) at a certain sky location, the pipeline assigns a running ID number to the sky coordinates of the detected object. Consequently, possible re-brightening of the source in the following nights is not given a new ID number and it is not considered among fast-transient candidates presented in this paper.

Our searches returned a large number of candidates ($>600,000$) most of which are spurious detections, transient/variable objects evolving at long timescales, or Galactic in origin.  We reduce the number of spurious detections by requiring candidates to be detected $\geq$2 consecutive times. Further selection criteria include constraints on the duration of the transients and the rejection of those likely of Galactic origin.   

We excluded those events detected at the beginning and/or at the end of each night, constraining the maximum evolution timescales within the time spent on a target field on each night, typically $\sim$\,1\,hr (Table\,\ref{tab:log}).  An example of a transient rejected from our sample because it was detected too close to the end of the night is provided in Figure\,\ref{fig: lc example ignored}. 

Searching separately for transients evolving at different timescales (e.g., 2 minutes against 20 minutes full-evolution time) brings several advantages.  For example, it enables the rejection of Galactic sources that emit rapid outbursts such as dwarf novae and some active M-dwarfs, favouring the identification of individual minute-timescale bursts. We enhanced the completeness of our searches by allowing the pipeline to `miss' a number of detections (termed `holes') between the first and the last detection of a candidate, on the night of first detection. We chose the ratio between the number of holes and the minimum number of detections to be $\sim$1/3.  Table\,\ref{tab:minmax detections} summarises the number of holes that we allowed to be present for each timescale, constrained between the minimum and maximum number of detections. Figure\,\ref{fig:selection criteria} helps visualise the selection criteria on the transient duration.

As we aim at identifying extragalactic fast transients, we introduced selection criteria to reduce the number of Galactic contaminants (variable and flare stars) in the sample. Using the Source Extractor \citep[\textsc{SExtractor},][]{Bertin2010} software on $g$-band stacks, we exclude those sources with a counterpart detected within a 2.2\,arcsec radius with a star/galaxy classification value \textsc{CLASS\_STAR}\,$> 0.95$. Such a threshold accommodates the change in point spread function (PSF) across the large field of view of DECam and is conservative because it is more likely that stars are classified as galaxies than vice-versa. \cite{Bleem2015} calculated that a \textsc{CLASS\_STAR} threshold of 0.95 is expected to include 94\% of all possible galaxies in the field and excludes 95\% of all stars.
First, we use \textsc{CLASS\_STAR} to reduce the number of bright false positives without significant probability of missing true extragalactic sources, other than quasi-stellar objects (QSOs).  Then, we use the \textsc{SPREAD\_MODEL} parameter (computationally more expensive than \textsc{CLASS\_STAR}) to improve the classification of those sources that survived our selection (Table~\ref{tab: shortlisted candidates}-\ref{tab: SExtractor class}).  The \textsc{SPREAD\_MODEL} value does not depend directly on the S/N of the source, so it can separate stars from galaxies more effectively than \textsc{CLASS\_STAR} close to the detection limit \citep[see for example][]{Sevilla-Noarbe2018}.  According to \cite{Annunziatella2013}, a threshold of 0.005 provides an optimal compromise between a reliable classification and a low contamination, with sources with \textsc{SPREAD\_MODEL} $< 0.005$ being likely stellar.

\begin{table}
\caption[Criteria used for the transient selection.]{Criteria used for the transient selection. Each row represents a search criterion adopted for each field for each observing night.  The first column presents the minimum number of times that a candidate was detected by the {\it Mary} pipeline; the second column presents the maximum number of images between the first and last detections of each candidate on a given night (extremes included); the third column shows the maximum number of non-detections (`holes') allowed between the first and last detection; the last column indicates the minimum number of `empty' images, both at the beginning and at the end of each given night, in which the candidate must {\it not} be detected in order to pass the selection. } 
\centering
	\begin{tabular}{cccc} 
		\hline\hline
		min det & max det & holes & empty \\
        \hline
		2 & 2 & 0 & 1 \\
		3 & 5 & 1 & 2 \\
		6 & 8 & 2 & 2 \\
		9 & 11 & 3 & 2 \\
		12 & 16 & 4 & 2 \\ 
		17 & 23 & 6 & 2 \\  
		24 & 32 & 8 & 2 \\ 
		33 & 47 & 12 & 2 \\  
		48 & 64 & 16 & 2 \\
		\hline
	\end{tabular}
	\label{tab:minmax detections}
\end{table}

\subsection{Stacking multiple images}
\label{subsec: searches in stacks}

\begin{figure}
\centering
	\includegraphics[width=\columnwidth]{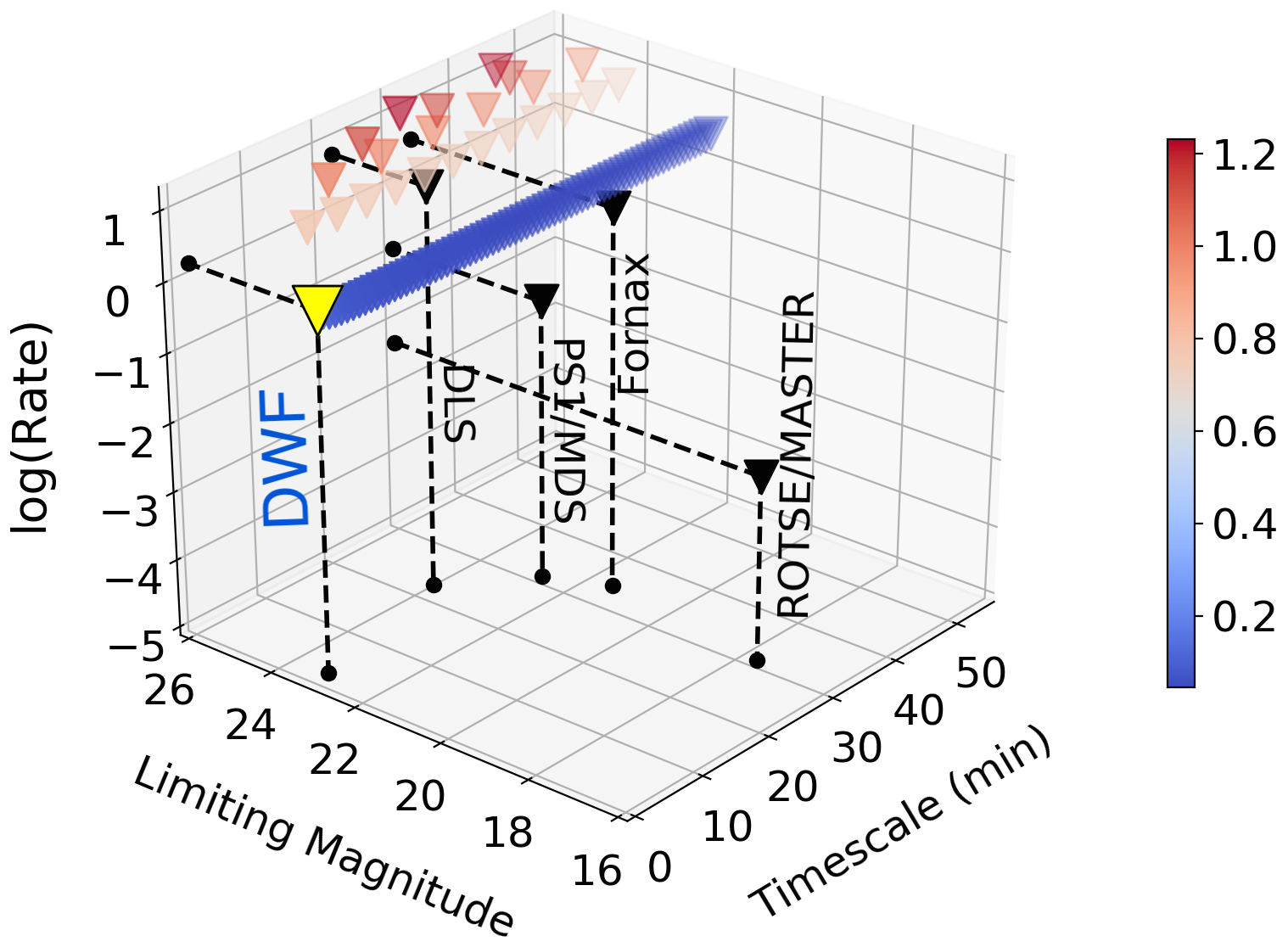}
    \caption[Plot of the new phase-space region explored.]{\small Plot of the new phase-space region explored, where rates of extragalactic fast transients (deg$^{-2}$day$^{-1}$) are plotted at each combination of limiting magnitude (i.e., depth) and timescale (expressed in minutes). The colour-map also represents the logarithm of the rates and helps visualising the differences between different points.  
    Triangles indicate upper limits placed during the systematic exploration of the DWF data set presented in this paper, assuming that all the eFTCs that passed our selection criteria are Galactic or spurious detections.
    The stacking of sets of images allows the exploration of different depth regimes, thus our results approximately describe a surface in the limiting~magnitude -- timescale -- transient\,rates phase-space.    The yellow triangle represents the shortest timescale that we can explore ($\tau=1.17$\,min), for which we obtain $R_{eFT} <  1.625$\,deg$^{-2}$day$^{-1}$.
    Black triangles indicate upper limits for past surveys, here presented as in \cite{Berger2013}. 
    }
    \label{fig:surveys}
\end{figure}

In Section~\ref{subsec: fast-cadenced images} we explained how we explored a dataset made of a large number of 20\,s exposures acquired with regular cadence. The exploration of the shortest timescale dictated by the cadence returns one point in the 3D space defined by timescale, depth, and areal rate \citep[see Figure\,\ref{fig:surveys} and][]{Berger2013}.  Assuming a constant limiting magnitude for each set of images, it is possible to search for transients exploring several timescales, potentially from less than the exposure time up to the duration of the observation of the target field, thus obtaining an array of areal rates that, if well-sampled, defines a broken line in Figure\,\ref{fig:surveys}.

We enrich the exploration of the parameter space of interest by dividing the images taken on each night in sets of 5, 9, 13, and 17 frames to be stacked together. The larger the number of images stacked together, the deeper we can search and therefore explore a bigger volume of Universe and detect fainter transients. This comes at the cost of reducing the temporal resolution and increasing the timescales of the transients to be discovered, losing information on their evolution, and reducing the effective areal exposure (see Section\,\ref{sec: rates}) given the finite total number of images available. The systematic exploration of pairs of timescale and limiting magnitudes defines a surface in Figure\,\ref{fig:surveys}.

The criteria to select eFTCs in series of stacked images are the same as described in Section\,\ref{subsec: selection criteria} and in Figure\,\ref{tab:minmax detections}, with the sole difference that the minimum number of `empty' stacked images, both at the beginning and at the end of each given night, in which the candidate must not be detected in order to pass the selection is always equal to 1. The choice of the number of images to stack (1, 5, 9, 13, 17) and the type of stacking (median) are dictated by technical reasons. Stacking more than 17 images together would cause the analysis to be meaningless in several cases, as no transient would possibly meet the criteria for the selection. We found that considering steps of 4 images balances well the need to densely sample the exploration space and the availability of computational resources, highly demanded when running the pipelines on thousands of images. Moreover, stacking an odd number of images is particularly suitable for median-stacking. The limitation to stack images considering median values derives from the structure of the {\it Mary} pipeline used to perform the analysis, which lacks a cosmic ray-rejection module that reduces the number of false positives when stacking a large number of images using averaging. In fact, average-stacking would have allowed us to be more sensitive to bright events with very short duration, thus we acknowledge that average-stacking would have been the preferred choice to adopt in this work.  We are planning to adopt average-stacking in future work. Nevertheless, median-stacking images allowed us to achieve excellent results.

 \begin{figure}
 \centering
	\includegraphics[width=\columnwidth]{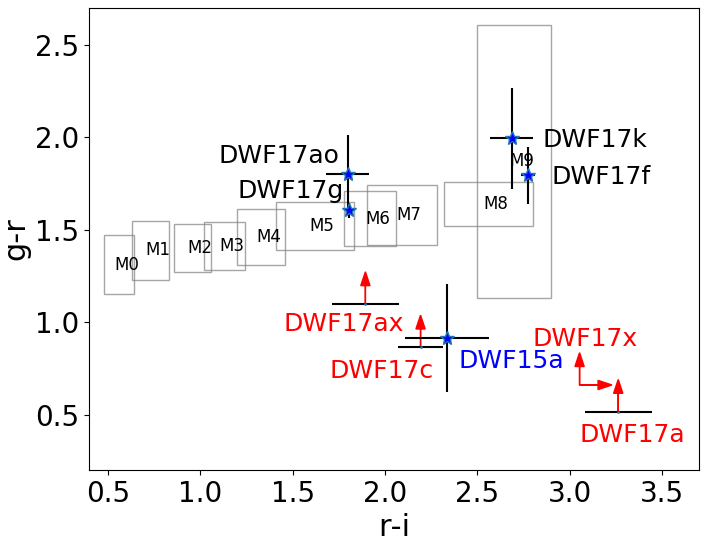}
    \caption[Colour-colour plot.]{Colour-colour plot showing measurements of possible counterparts to the eFTCs listed in Table\,\ref{tab: shortlisted candidates}. Magnitude values were calibrated using the APASS catalogue. Boxes frame regions of the colour-colour plot where different types of M-dwarfs lie  \citep{West2011}. 
    }
    \label{fig:grri}
\end{figure}

\section{Results}
\label{sec: results}

In order to carry out the searches described in Section\,\ref{sec: search for optical bursts}, we ran the {\it Mary} pipeline 2744 times in total, each time processing 59 CCDs in parallel.  The analysis of our dataset returned 318\,672 candidates\footnote{This number includes possible repetitions of the same candidate when stacking different sets of multiple images.}, including thousands of real variable and transient sources along with a large number of false positives. We reduced the number of candidates to $\sim 10,000$ by applying the selection criteria on the duration of the transients, presented in Table\,\ref{tab:minmax detections}. All candidates were visually inspected at this preliminary stage and further inspection was performed after applying the following cuts.  

When all the selection process described in Section\,\ref{subsec: selection criteria} was completed, 1846 candidates remained. Visual inspection of those candidates, exclusion of asteroids, and the removal of repetitions due to the detection of the same object when stacking different sets of images left us with 25 candidates. Of those 25 candidates, we classified one as AGN activity, 2 were already catalogued as variable stars in the Vizier database\footnote{\url{https://vizier.u-strasbg.fr/viz-bin/VizieR}}, and 13 have parallax or high proper-motion measurements reported in the second {\it Gaia} Data Release \citep{Gaia2016,Gaia2018DR2}.  Excluding those Galactic and nuclear sources from our sample, 9 eFTCs constitute our short list.  Their identifications and sky coordinates are presented in Table\,\ref{tab: shortlisted candidates}.    
We stress that many more real transient sources were identified than those reported in this paper, however a limited number of those passed the selection criteria that we established.

\begin{table*}
\caption[fast-transient candidates that passed our selection criteria.]{Identification name, target field, and coordinates of those fast-transient candidates that passed our selection criteria. 
 The magnitude variation is computed as the difference between the deepest $g$-band measurement available and the observed peak magnitude, therefore the actual variation in magnitude is likely larger than what is reported here. The light curves of the eFTC presented here are shown in Figure\,\ref{fig: lc example DWF17ax} and in Figures\,\ref{fig:lc eFTC stellar}-\ref{fig:lc eFTC non-stellar}. Candidates in the lower half of the table, below the horizontal line, are those for which \textsc{CLASS\_STAR} alone on the multi-band data cannot determine whether the quiescent counterpart to the transients is stellar. 
 } 
\centering
	\begin{tabular}{ccccccccc} 
		\hline\hline
		ID & Field & RA & Dec & --$\Delta${\it g} &$g$&$r$&$i$&$z$\\
        \hline
DWF15a& 4hr&  4:11:05.702& $-$55:40:17.87  & $\sim 5.0$ & $24.78 \pm 0.20$ & $23.86 \pm 0.21$ & $21.53 \pm 0.08$ & - \\
DWF17a & Antlia&  10:25:53.642& $-$35:31:20.67 &$>$2.4 & $>$24.15 & $23.64 \pm  0.18$ & $20.37 \pm 0.01$ & $19.90 \pm 0.023$ \\
DWF17c& Antlia&10:28:48.603 & $-$36:07:00.54 & $>$2.4 & $>$24.15& $23.29 \pm 0.12$ & $21.09 \pm 0.02$ & $20.97 \pm 0.07$ \\
DWF17f & Antlia& 10:27:36.287& $-$35:36:59.79 & 1.8 & $23.53 \pm 0.15$ &$21.74 \pm 0.04$ & $18.96 \pm 0.01$ & $18.97 \pm 0.01$\\
DWF17g& Antlia& 10:25:47.560& $-$35:41:54.92 &0.8& $21.61 \pm 0.04$ & $20.00 \pm 0.01$ & $18.198 \pm 0.003$ & $17.990 \pm 0.005$ \\
\hline
DWF17k& FRB131104&  6:47:05.788& $-$51:27:38.88 &$\sim$4.8 &$25.19 \pm 0.25$ &$23.19 \pm 0.11$ &$20.50 \pm 0.07$ & - \\
DWF17x& FRB131104& 6:45:04.601& $-$51:38:18.26 & $>$4.6 & $>$25.0 & $>$24.3 & $21.29 \pm 0.26$ &-\\
DWF17ao & Prime& 5:52:29.591& $-$60:49:50.92 & 3.6 & $25.46 \pm 0.18$ & $23.66 \pm 0.10$ & $21.86 \pm 0.05$ & $21.16 \pm 0.01$ \\
DWF17ax& Prime&  5:59:00.662& $-$62:02:11.03 & $>$5.6 & $>$25.6 & $24.53 \pm 0.16$& $22.64 \pm 0.09$ &$21.98 \pm 0.02$\\
\hline
	\end{tabular}
	\label{tab: shortlisted candidates}
\end{table*}


\begin{table*}
\caption[]{\textsc{SExtractor} star/galaxy classification of short-listed eFTCs in quiescence in deep images using the  \textsc{CLASS\_STAR} (C\_S) and the \textsc{SPREAD\_MODEL} (S\_M) parameters. 
The last column indicates whether the sources were classified as stellar with score S/G $>0.95$ in at least one band. In such cases, we consider the \textsc{SExtractor} classification to be in support of the eFTCs being Galactic stellar flares.   } 
\centering
	\begin{tabular}{cccccccccc} 
		\hline\hline
ID & C\_S & S\_M &  C\_S & S\_M &  C\_S & S\_M &  C\_S & S\_M & S \\
 & $g$ & $g$ & $r$ & $r$ & $i$ & $i$ & $z$ & $z$ & \\ 
\hline
DWF15a & 0.59 & $0.0001 \pm 0.0093$  & 0.22 & $-0.0019 \pm 0.0054$ & 0.97 & $-0.0012 \pm 0.0023$ & None & None & Y\\ 
DWF17a & None & None  & 0.01 & $0.0069 \pm 0.0055$ & 0.98 & $-0.0002 \pm 0.0005$ & 0.97 & $-0.0000 \pm 0.0006$ & Y\\ 
DWF17c & None & None  & 0.74 & $-0.0027 \pm 0.0038$ & 0.98 & $0.0002 \pm 0.0009$ & 0.96 & $-0.0004 \pm 0.0014$ & Y\\ 
DWF17f & 0.55 & $-0.0095 \pm 0.0046$  & 0.98 & $0.0001 \pm 0.0010$ & 0.99 & $-0.0002 \pm 0.0002$ & 0.99 & $-0.0012 \pm 0.0003$ & Y\\ 
DWF17g & 0.72 & $0.0032 \pm 0.0011$  & 0.98 & $0.0039 \pm 0.0003$ & 0.98 & $0.0050 \pm 0.0001$ & 0.98 & $0.0052 \pm 0.0002$ & Y\\ 
DWF17k & 0.04 & None  & 0.00 & $-0.0017 \pm 0.0033$ & 0.04 & $0.0006 \pm 0.0009$ & None & None & Y\\ 
DWF17x & None & None  & None & None & 0.36 & $-0.0092 \pm 0.0061$ & None & None & Y\\ 
DWF17ao & 0.00 & None  & 0.83 & $-0.0043 \pm 0.0037$ & 0.67 & $0.0005 \pm 0.0006$ & 0.85 & $-0.0002 \pm 0.0004$ & Y\\ 
DWF17ax & None & None  & 0.00 & $0.0038 \pm 0.0083$ & 0.00 & $0.0003 \pm 0.0012$ & 0.33 & $-0.0007 \pm 0.0006$ & Y\\ 
\hline
	\end{tabular}
	\label{tab: SExtractor class}
\end{table*}

\begin{table*}
\caption[Searches for multi-wavelength counterparts.]{This table indicates when gamma-ray and radio telescopes were observing the region of sky where a subset of eFTCs (Table\,\ref{tab: shortlisted candidates}) were discovered. The onset time indicated represents the MJD of the last non-detection of the eFTC. We looked into {\it Swift}/BAT data to identify gamma-ray counterparts, and we used the Parkes and Molonglo radio telescopes to search for FRBs. We coordinated {\it Swift}, Parkes, and Molonglo to observe fields simultaneously with DECam as part of the DWF programme. In addition, we looked for gamma-ray triggers issued by the {\it Fermi} satellite $\pm 1$\,day from the onset time, with the caveat that the location where the eFTC was detected may have been outside the {\it Fermi} field of view for part of the time.   } 
\centering
	\begin{tabular}{cccccccc} 
		\hline\hline
		ID & Onset & {\it Fermi} & {\it Swift}/BAT & Parkes & Molonglo & Gamma-ray & FRB \\
		  & (MJD) & (days) &(minutes)&(minutes)&(minutes)& detections & detections\\
        \hline
DWF17k &57789.27229 & +/- 1 & +8 +30 & --5 +60 & --8 +65 & 0 & 0\\
DWF17x &57786.29119 & +/- 1 & --10 --3; +3 +27 & --25 +45 & - & 0 &0 \\
DWF17ao &57790.24767 & +/- 1 & - & --45 +35 & --2 + 25 & 0 &0 \\
DWF17ax &57789.24627 & +/- 1 & --30 --18& --20 +30 & --47 + 12 & 0 &0 \\
\hline
	\end{tabular}
	\label{tab: multi-wavelength}
\end{table*}

 \begin{figure*}
 \centering
 	\includegraphics[width=0.86\columnwidth]{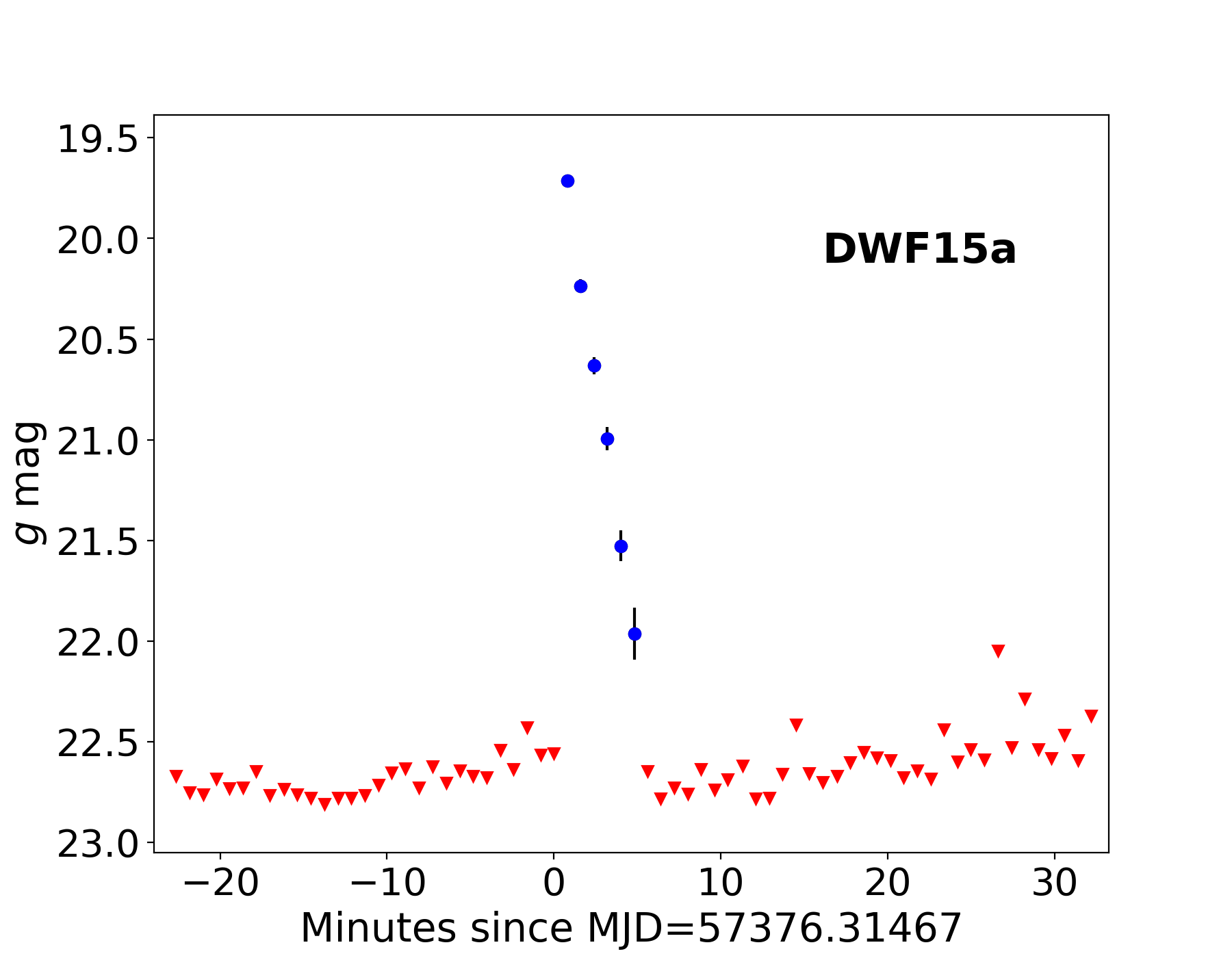}
 	\includegraphics[width=0.86\columnwidth]{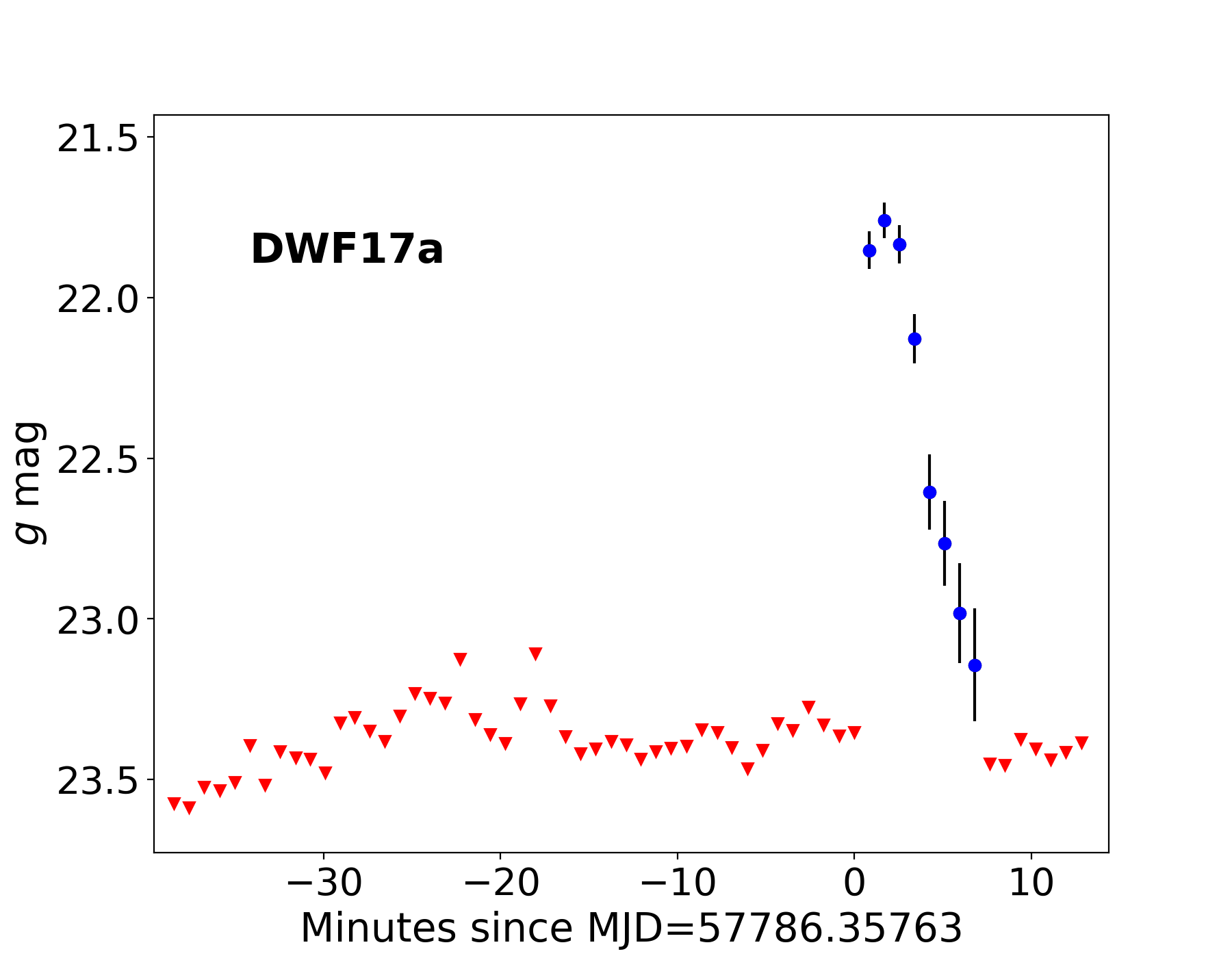}
	\includegraphics[width=0.86\columnwidth]{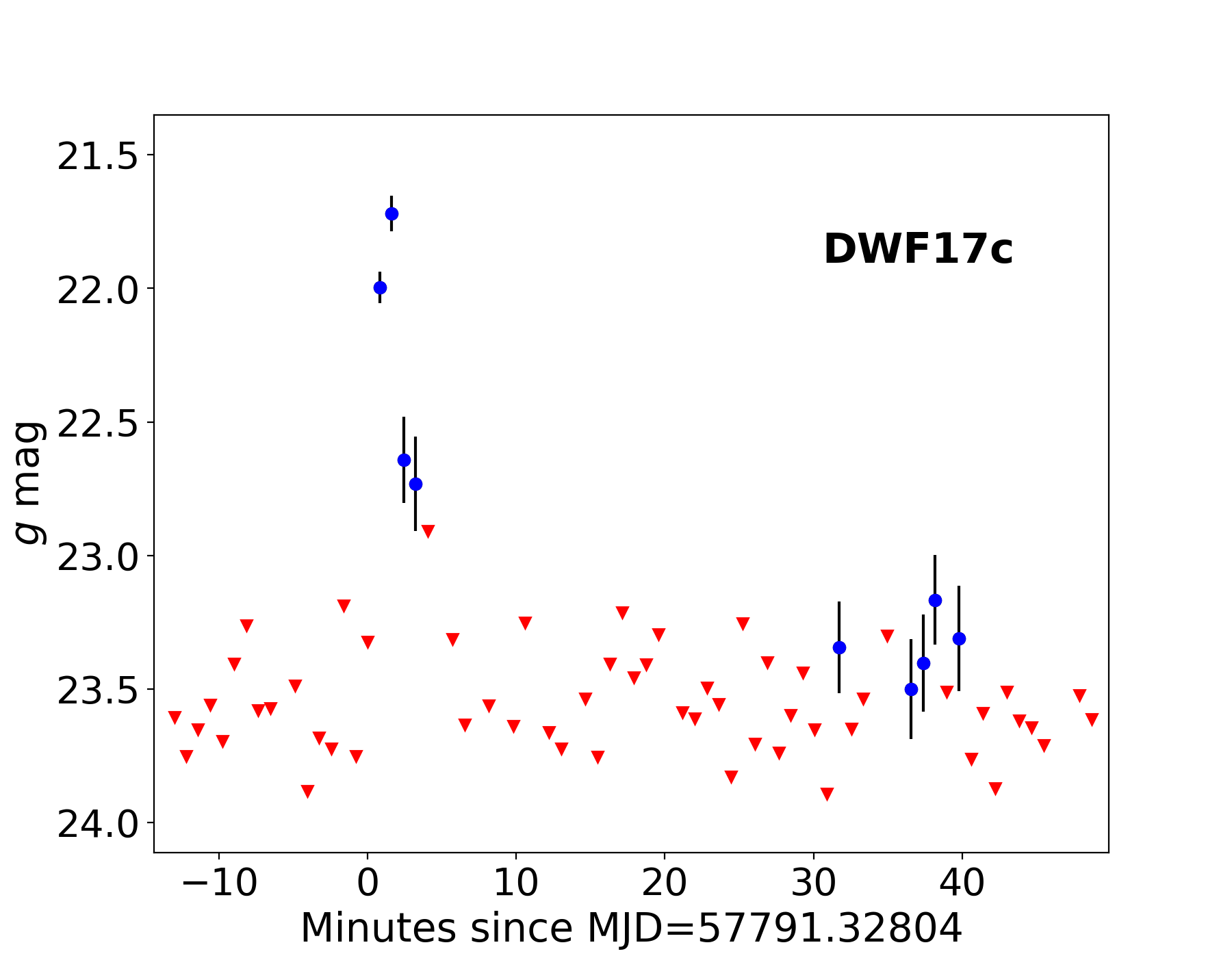}
	\includegraphics[width=0.86\columnwidth]{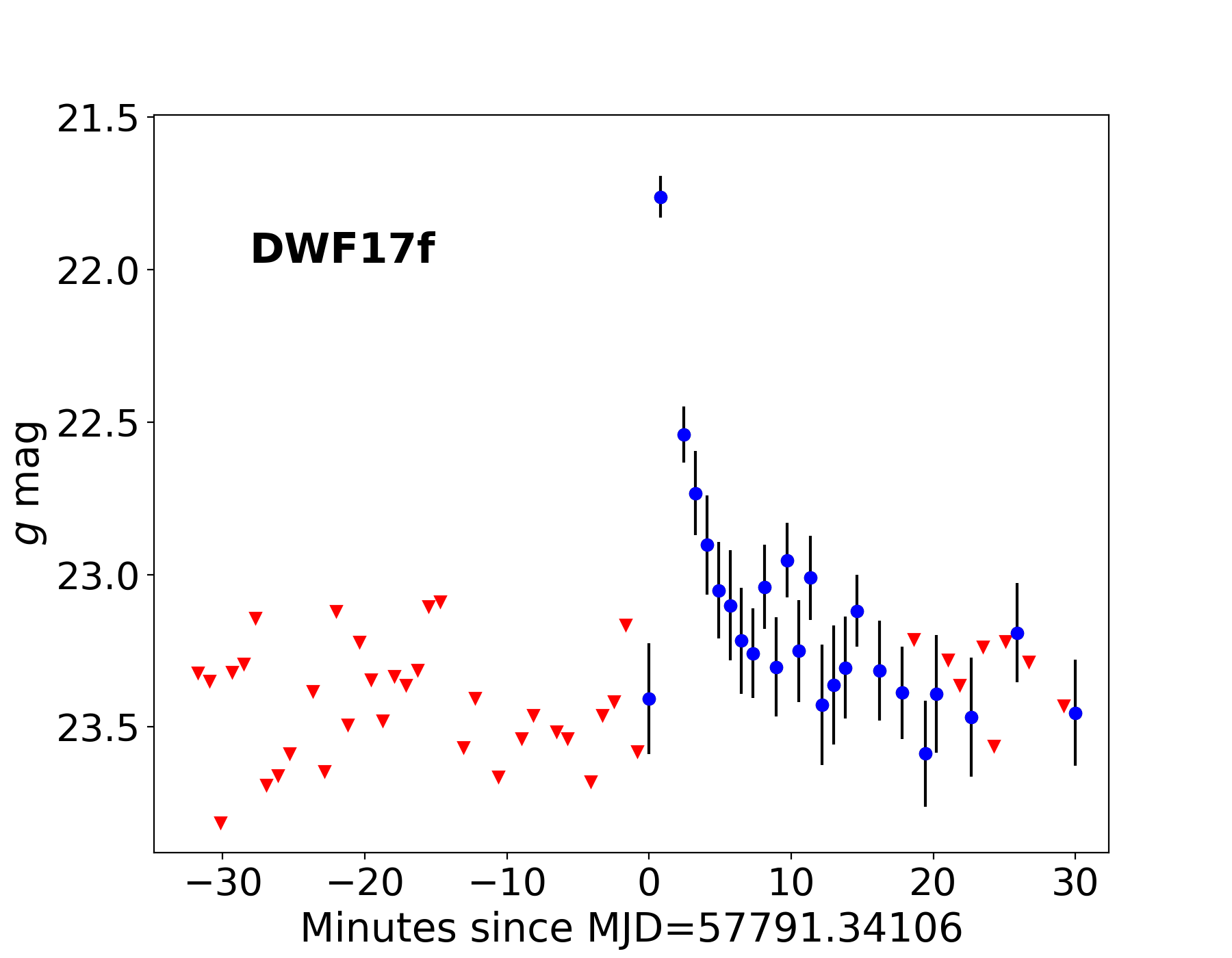}
	\includegraphics[width=0.86\columnwidth]{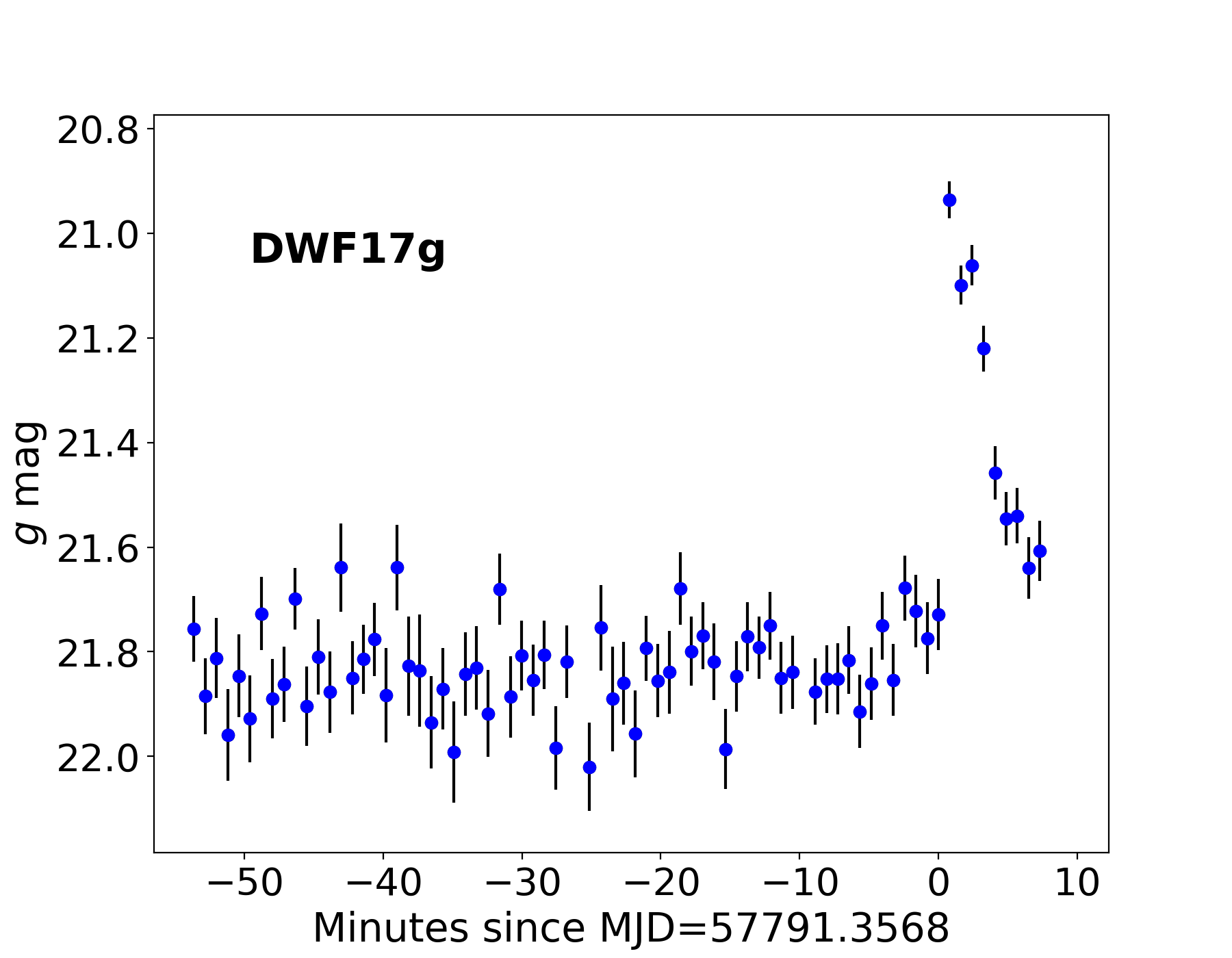}
    \caption[Light curves of the selected fast-transient candidates with stellar counterparts.]{Light curves of selected extragalactic fast-transient candidates (eFTCs) that we classify as stellar flares based on the multi-band S/G separation with both \textsc{CLASS\_STAR} and \textsc{SPREAD\_MODEL} (Table\,\ref{tab: SExtractor class}).  Red triangles represent 5$\sigma$ forced-photometry upper limits.  Candidates DWF17f and DWF17g show measurements on the last epochs of the observing sequence, missed by the image-subtraction pipeline with threshold at $\sim 7 \sigma$ significance. Details about these sources are presented in Table\,\ref{tab: shortlisted candidates}.  }
    \label{fig:lc eFTC stellar}
\end{figure*}

 \begin{figure*}
 \centering
	\includegraphics[width=0.86\columnwidth]{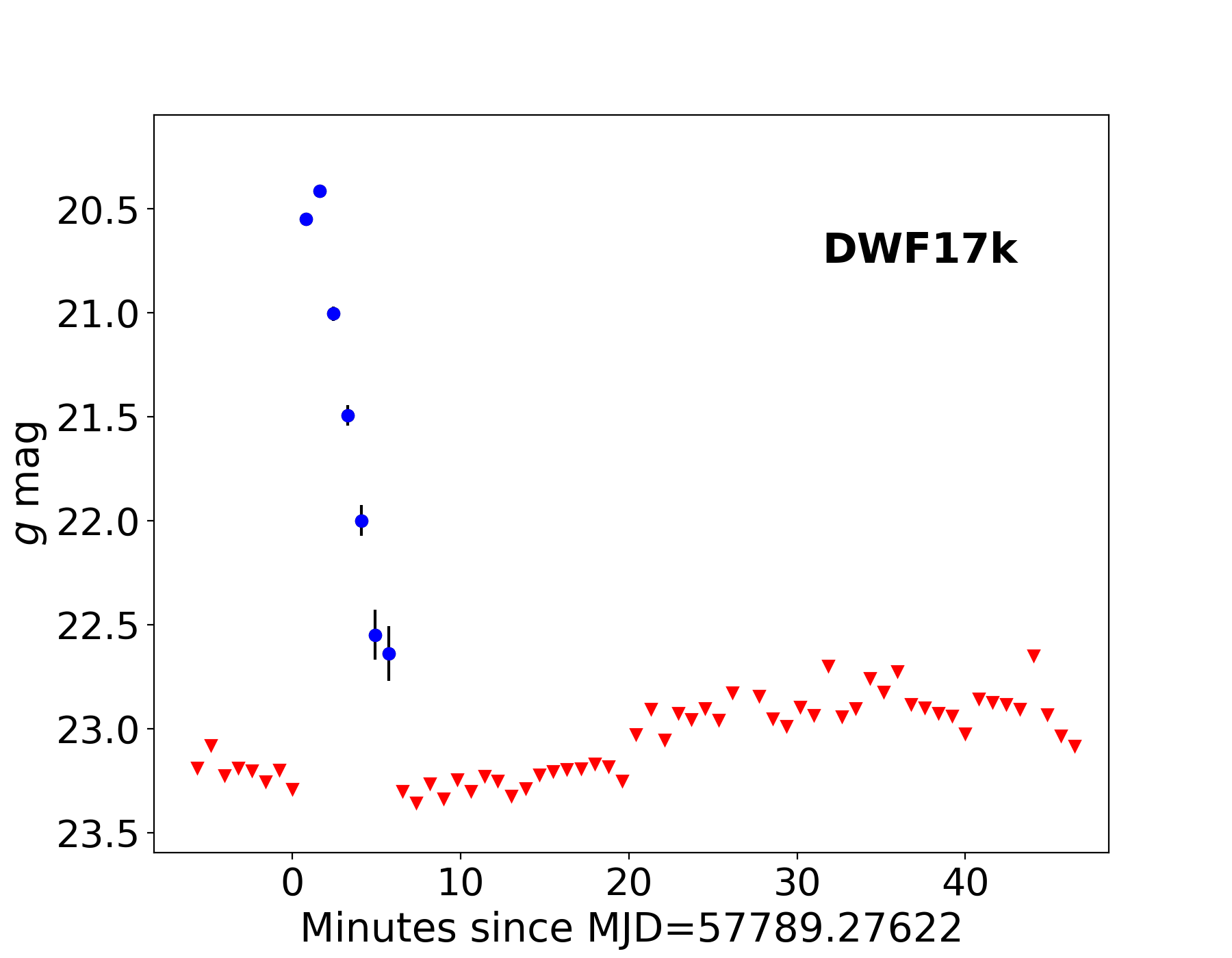}
	\includegraphics[width=0.86\columnwidth]{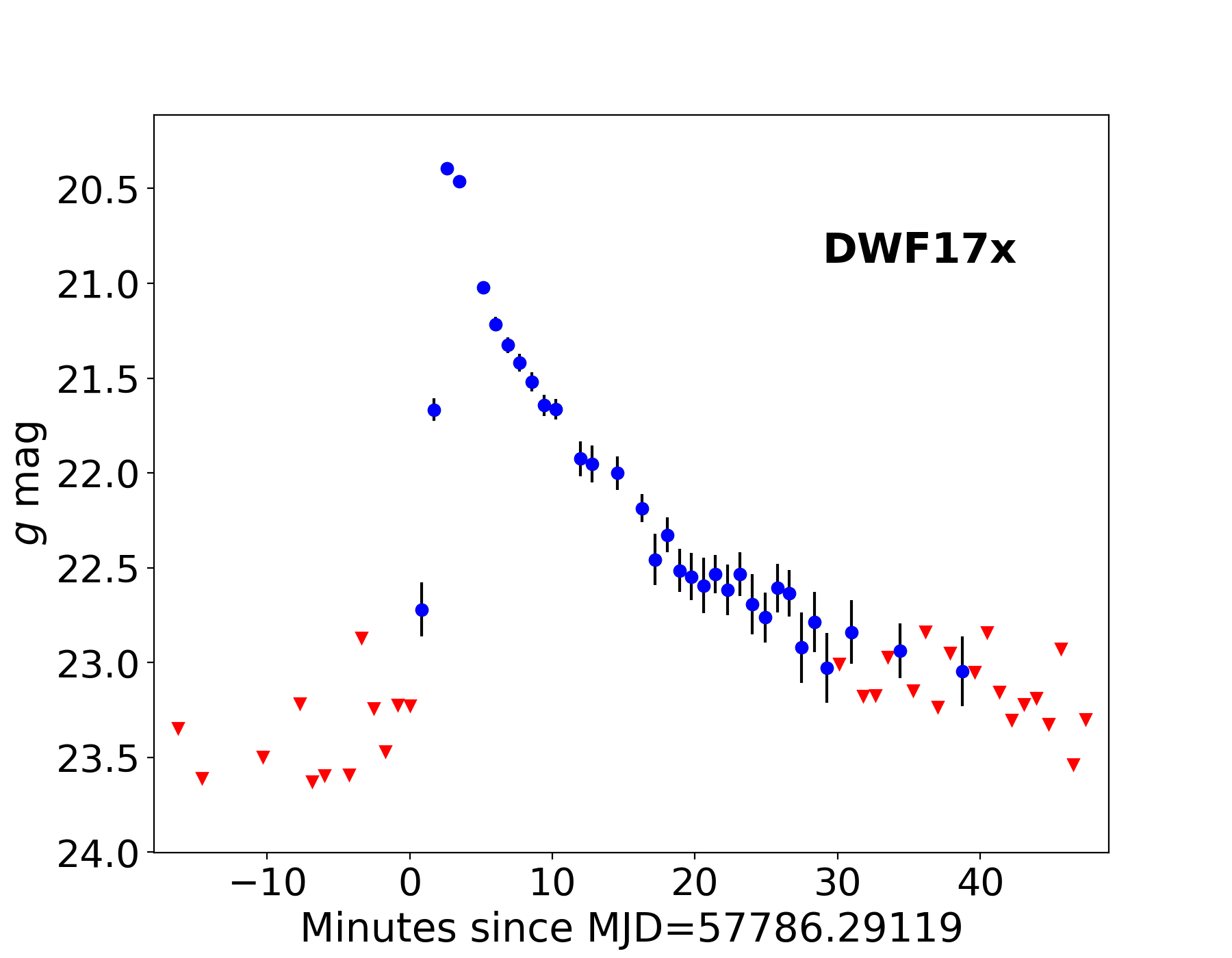}
	\includegraphics[width=0.86\columnwidth]{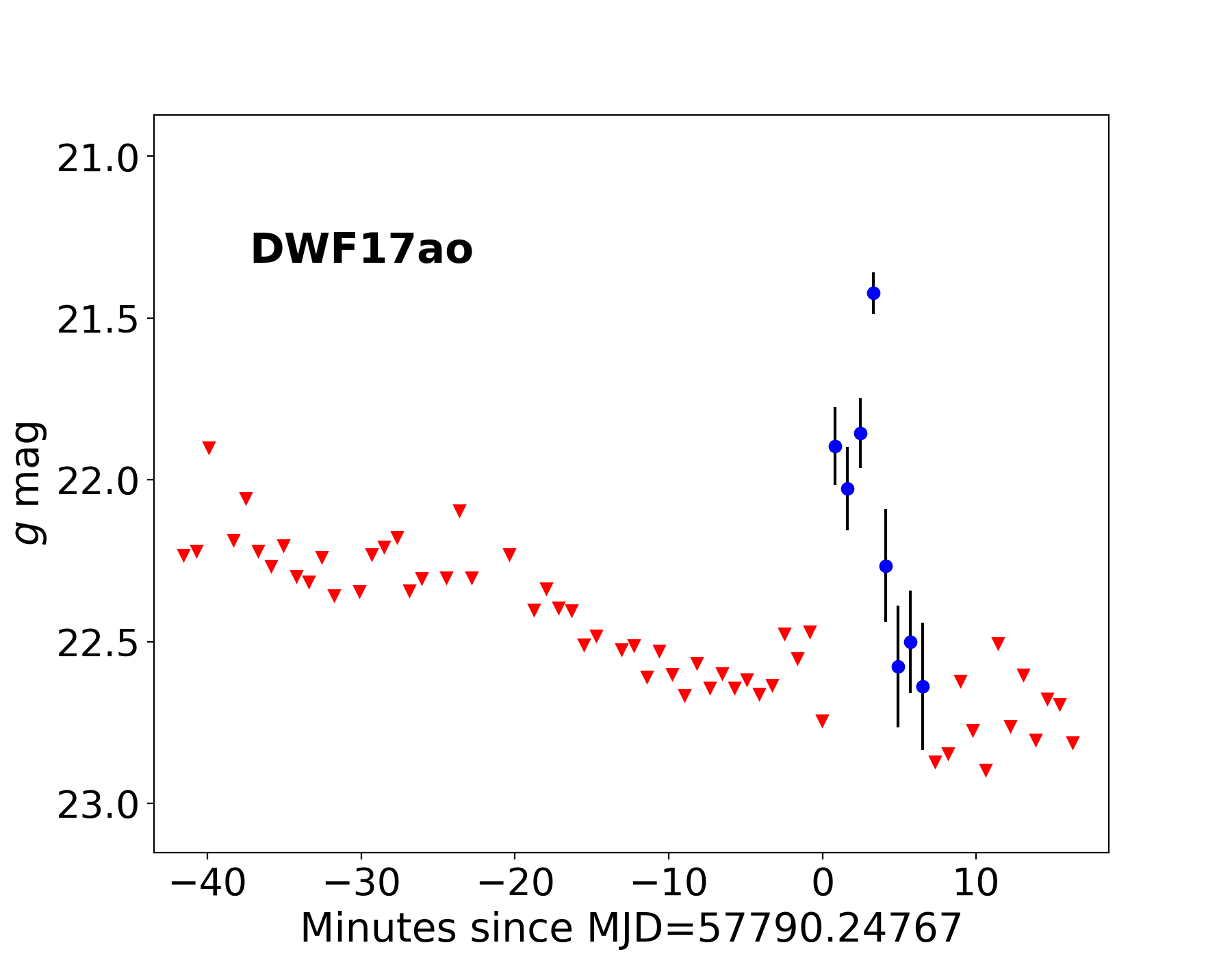}
	\includegraphics[width=0.86\columnwidth]{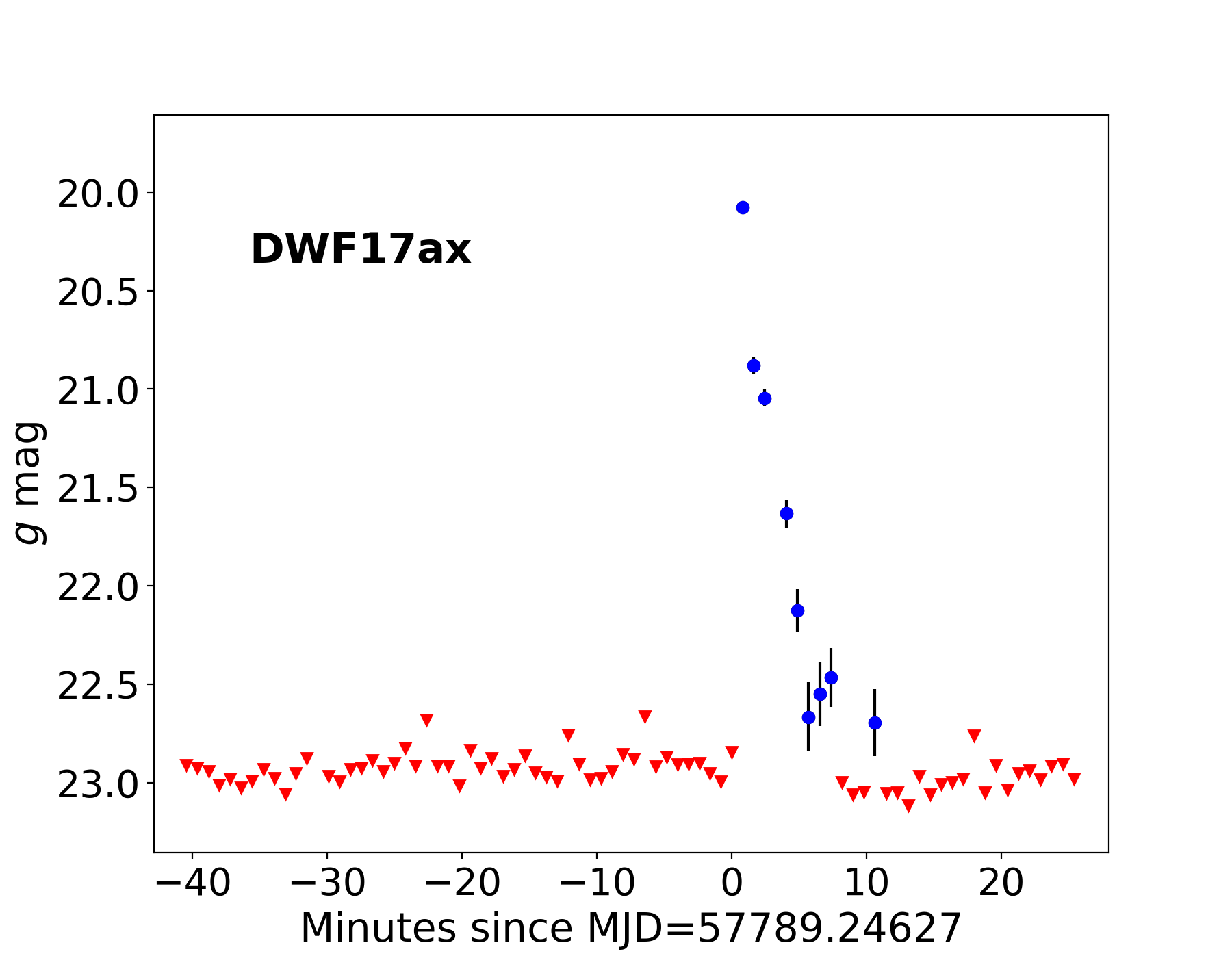}
    \caption[Light curves of the selected fast-transient candidates.]{Light curves of selected extragalactic fast-transient candidates (eFTCs) that we cannot classify as stellar flares based on \textsc{CLASS\_STAR}, but that likely have stellar progenitors based on \textsc{SPREAD\_MODEL} information.  Details about the eFTCs are presented in Table\,\ref{tab: shortlisted candidates}.  The light curve of DWF17ax can be found also in Figure\,\ref{fig: lc example DWF17ax}, where we show the effect of image stacking on a transient light curve.  }
    \label{fig:lc eFTC non-stellar}
\end{figure*}

\begin{figure}
 \centering
    \includegraphics[width=\columnwidth]{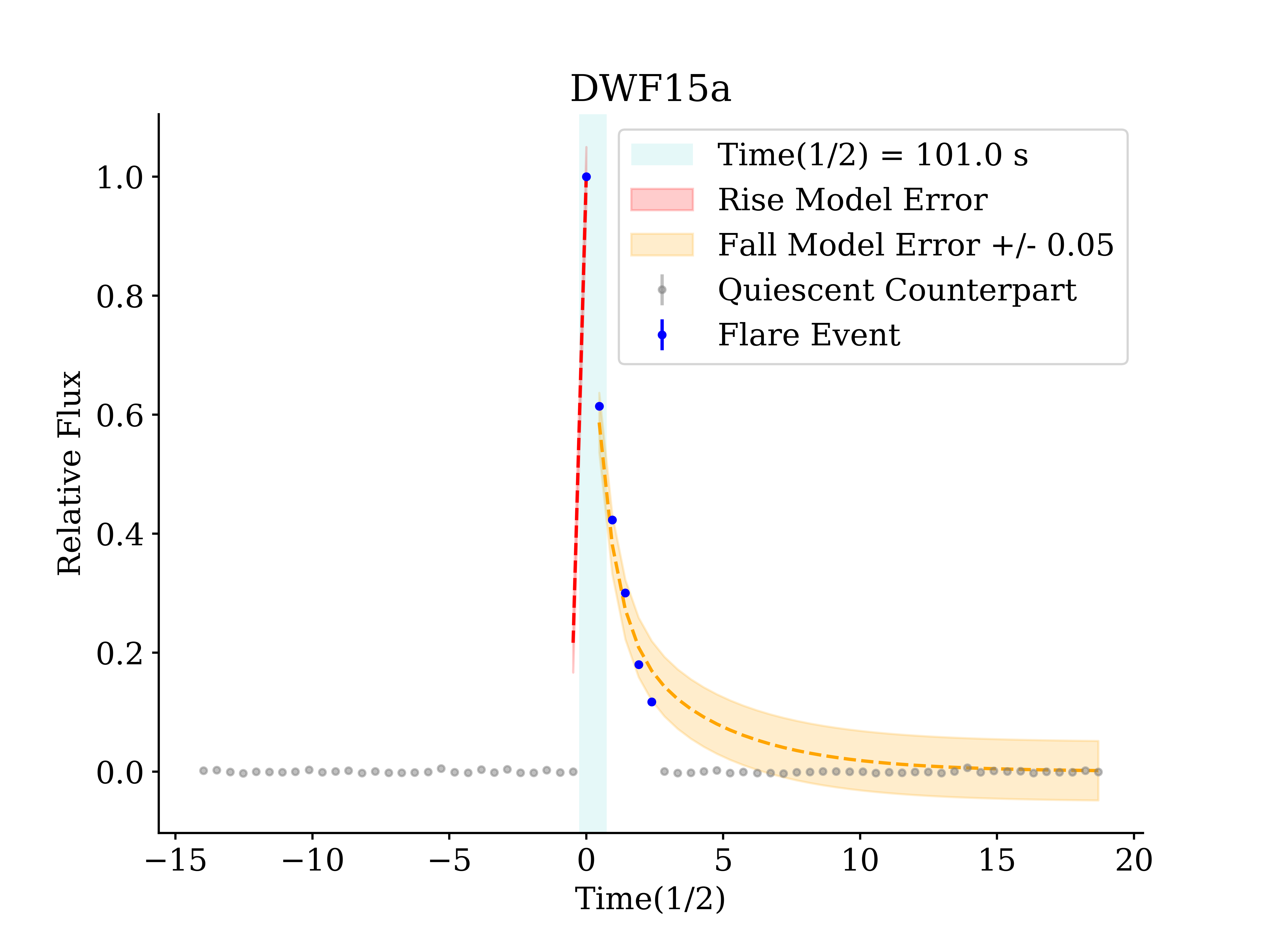}
	\includegraphics[width=\columnwidth]{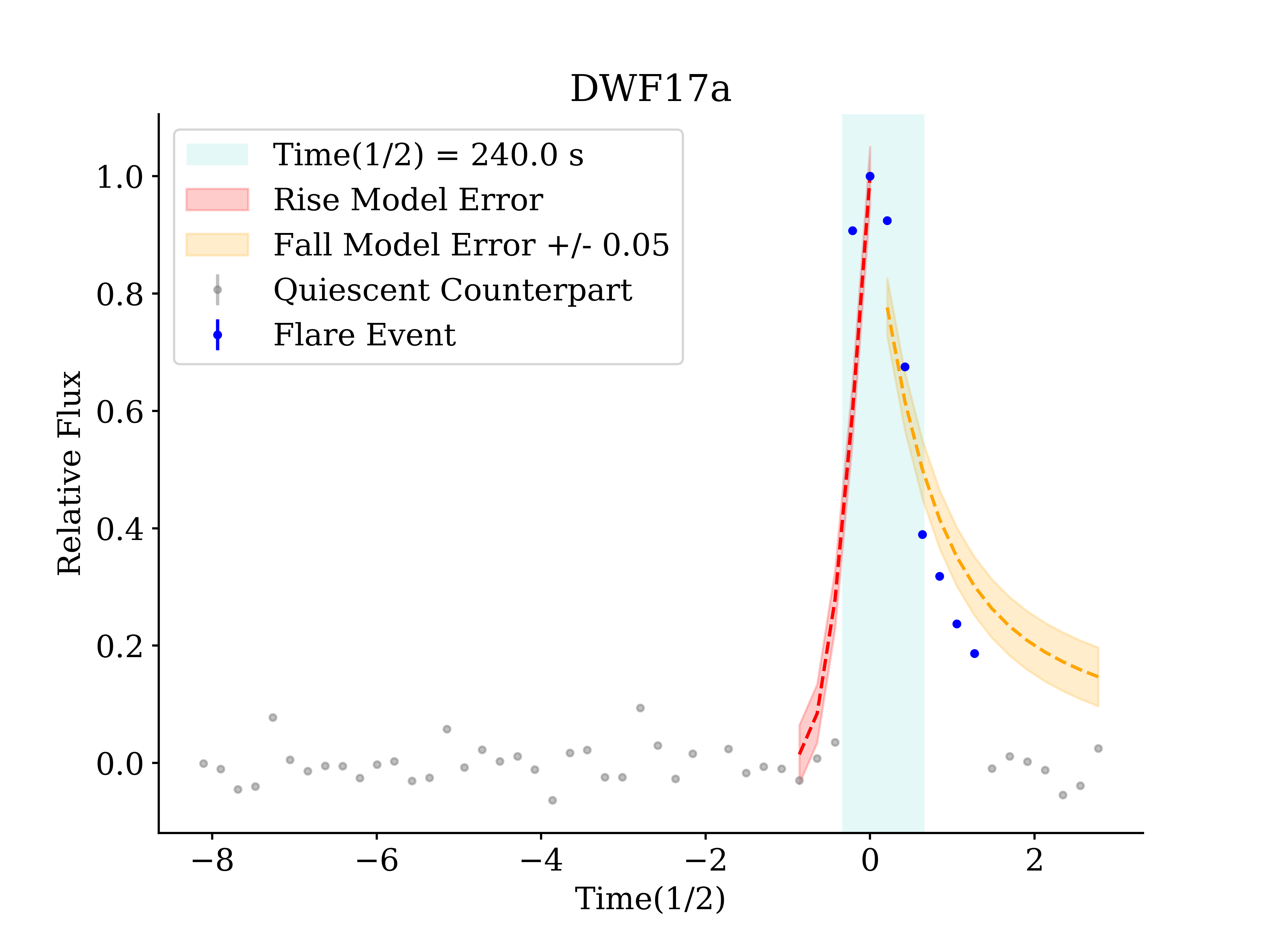}
	\includegraphics[width=\columnwidth]{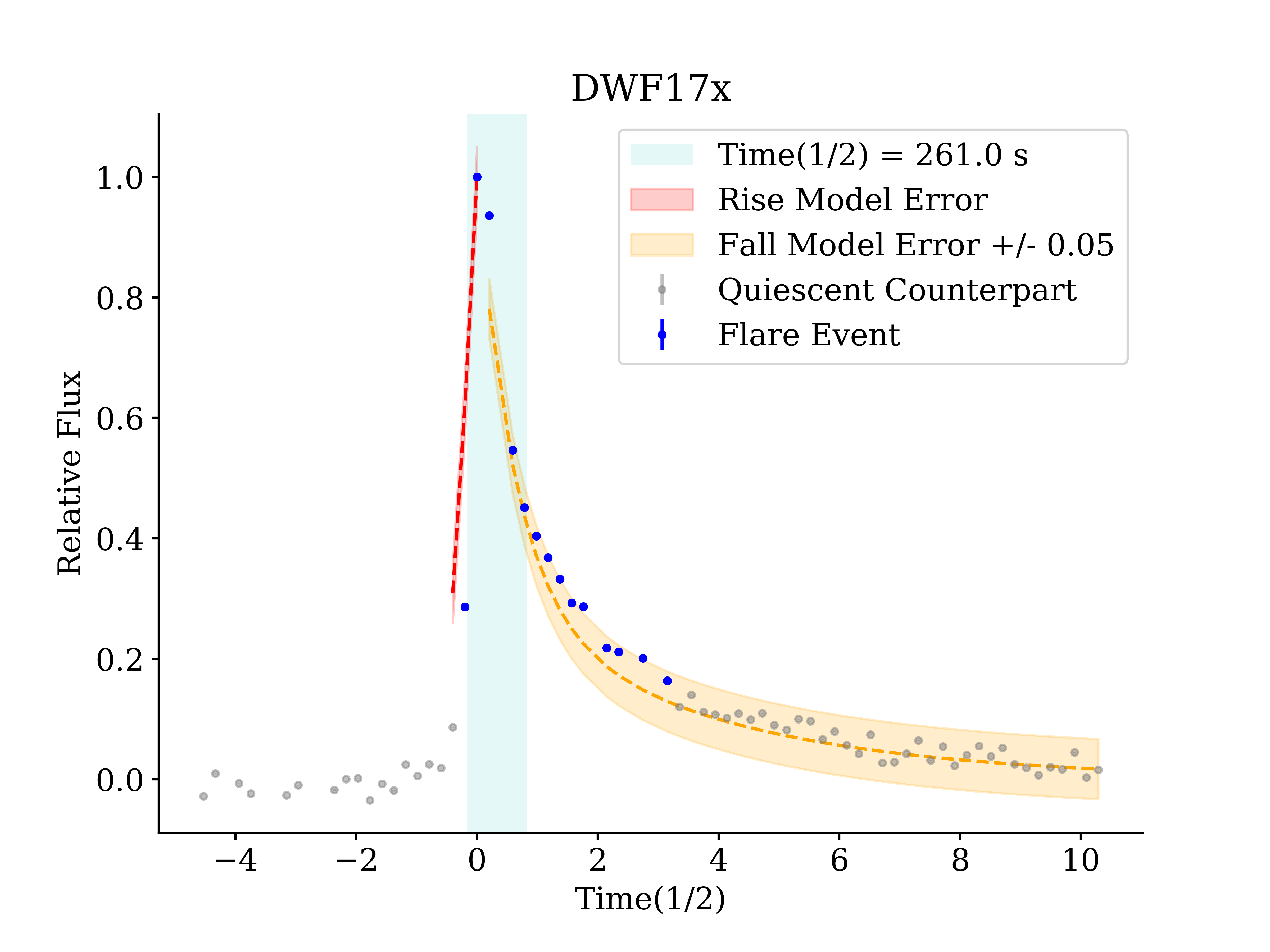}
    \caption[]{Flare-star model for the eFTC DWF15a (top), DWF17a (centre), and DWF17x (bottom).  Data are compared with template white light flares based on {\it Kepler} observations \citep{Davenport2014}, where Time(1/2) is the FWHM of the flare.  When no permanent source is detectable in fast-cadenced $g$-band images outside the flare times, we simulate a Gaussian distribution of data points centered on the magnitude value (or upper limit) calculated on deep stacks (see Table\,\ref{tab: shortlisted candidates}). We normalised the flux to the brightest point of the light curve, however the peak of the flare is likely brighter and shifted in time by $0 \lesssim t_{\rm peak} \lesssim 30$\,s, given the cadence of our observations. A better guess of time and flux of the peak would likely result in a better fit of the model to the light curves in some cases, but poorer in others.  }
    \label{fig: flare model DWF15a DWF17a DWF17x}
\end{figure}

We further investigate the colour information and behaviour of the light curves of the nine short-listed eFTCs using \textsc{SExtractor} \textsc{CLASS\_STAR} and \textsc{SPREAD\_MODEL} on deep stacks in $riz$ bands (where available).
Results of the multi-band S/G classification are presented in Table\,\ref{tab: SExtractor class}.  Five eFTCs (DWF15a, DWF17a, DWF17c, DWF17f, and DWF17g) can be classified as stellar if we consider again a S/G threshold \textsc{CLASS\_STAR}\,$> 0.95$. All 9 sources are likely stellar considering the \textsc{SPREAD\_MODEL} parameter, since \textsc{SPREAD\_MODEL}\,$< 0.005$ in at least one band in all cases. 
Colours are obtained with photometric measurements on deep stacks of images acquired before the transient detection.  Magnitudes are calibrated using the AAVSO Photometric All Sky Survey \citep[APASS,][]{Henden2016} catalogue.  Sources with a detectable counterpart in deep stacks and in different filters are plotted in Figure\,\ref{fig:grri} (see also Table\,\ref{tab: shortlisted candidates}).  
We also attempt a comparison between our data with flare star models computed from {\it Kepler} observations \citep{Davenport2014}.  One example of such a comparison is presented in Figure\,\ref{fig: flare model DWF15a DWF17a DWF17x} for the eFTC candidate DWF17a. The selected eFTCs are individually discussed below. As mentioned above, all these candidates are likely stellar based on the multi-band S/G classification.

\begin{itemize}
    \item DWF15a -- 
    Faint detections of the source in deep stacks place it outside the M-dwarf stripe in the $g-r, r-i$ colour plot. The light curve is consistent with a template flare star light curve (Figure\,\ref{fig: flare model DWF15a DWF17a DWF17x}).  
    \item DWF17a -- 
    The light curve evolves faster than the stellar flare model \citep{Davenport2014} shown in (Figure\,\ref{fig: flare model DWF15a DWF17a DWF17x})  We note that a large difference $r-i >3$ suggests the flare to be generated from a star of type later than M9.   
    \item DWF17c, DWF17f -- 
    The light curves compare well with the stellar flare model.  These sources can be classified as M7 (DWF17c) and M8-M9 (DWF17f) stellar flares.
    \item DWF17g -- Very likely M5-M6 star flare.  
    \item DWF17k -- 
    The location in the colour-colour plot suggests DWF17k could be a M8-M9 star flare.  A Galactic nature of this source is also supported by a good resemblance between the light curve and the template stellar flare we consider. 
    \item DWF17x -- 
    The upper limits in the colour-colour plot in Figure\,\ref{fig:grri} give little information about the nature of DWF17x.  The light curve matches well the flare-star model (Figure\,\ref{fig: flare model DWF15a DWF17a DWF17x}).  The small discrepancy could be due to the wrong choice of the peak time, that a higher cadence would have improved. 
    \item DWF17ao -- 
    The multi-peak light curve advocates for a flare star event that, according to the colour-colour plot in Figure\,\ref{fig:grri}, may have a M5-M6 star progenitor.  
    \item DWF17ax -- 
    The fast-cadence light curve of DWF17ax (Figure\,\ref{fig: lc example DWF17ax}, top-left panel) shows a possible multi-peak structure, common among flare stars.  If the transient is indeed a flare star, its progenitor's class could range between M5 and M7. 
\end{itemize}


\section{Searches for multi-wavelength counterparts}
\label{sec: multi-wavelength}
The DWF programme coordinates simultaneous multi-wavelength observations of the target fields.  Moreover, we take templates using the wide-field SkyMapper
telescope weeks before DWF observing runs and we use it to perform interleaved, nightly observations during DWF, and late-time regularly-cadenced observations weeks after DWF to characterise long-duration transients.  In this section we present searches for gamma-ray and radio signals possibly associated with the eFTCs that we selected using the Parkes, Molonglo, {\it Swift}, and {\it Fermi} observaotries. While large datasets were analysed, particular attention was given to those eFTCs for which the S/G separation is less robust because it relies on the \textsc{SPREAD\_MODEL} parameter only, instead of on both \textsc{SPREAD\_MODEL} and \textsc{CLASS\_STAR} parameters. Details and results of these searches are summarised in Table\,\ref{tab: multi-wavelength}. Searches for long-duration optical transient counterparts with SkyMapper are described in Section\,\ref{subsec: skymapper}.

\subsection{Searches for gamma-ray signals}
We explore data acquired with {\it Fermi} and {\it Swift} gamma-ray telescopes at times close to the last non-detection of our selected eFTC.

\subsubsection*{{\it Neil Gehrels Swift Observatory}}

Observations with the {\it Neil Gehrels Swift Observatory} (hereafter {\it Swift}) were performed under approved Cycle 11 and Cycle 13 programmes (PI Pritchard).  Moreover, the large FoV of {\it Swift}/Burst Alert Telescope \citep[BAT,][]{Barthelmy2005a} allowed DWF target fields to be observed even when the satellite was pointing at other scientific programme targets with its narrower-field instruments X-ray Telescope (XRT) and UltraViolet/Optical Telescope (UVOT). The {\it Swift} team smartly scheduled {\it Swift} observations of other science programmes to maximise the observability of DWF fields with BAT.   
We searched for gamma-ray counterparts to five most promising eFTCs in particular: DWF15a, DWF17k, DWF17x, DWF17ao, DWF17ax.  Some of the DWF {\it Swift} time had BAT, XRT, and UVOT on the target fields.  However, none of the sources were located within the FoV of the XRT and UVOT, therefore we limit our analysis to the BAT data. 

We considered the last optical non-detection as onset time (T0).  Onset times are listed in Table\,\ref{tab: multi-wavelength}.  We expect T0 to be accurate within 30\,s from the actual onset of the eFTCs because of the short rise-time of the transients and the high cadence of our observations. The nature of the eFTCs being uncertain, we explored a conservative time window of $\pm 1800$\,s around T0, larger than the evolution timescale of the candidates. Except DWF17ao, all candidates were located in the BAT FoV for some period of time within the search interval. DWF17ao occurred when {\it Swift} was in safehold, so no data were available. A wider temporal search (years before and after the events) is planned for future work.

When trying to identify gamma-ray counterparts, unconstrained by physical models, we searched:

\begin{itemize}
    \item The raw light curves, to see if there are any obvious GRB-like structures around T0.  We did not find GRB-like events in raw light curves.
    
    \item The available event data, from which we can make a background-subtracted (i.e., mask-weighted) light curve using the source location. In particular, we looked for astrophysical signals in mask-weighted light curves (e.g., burst-like features or some continuous time bins with count rate above $\sim 3\sigma$).  We also created images of the available event data interval and search for any detections at the source location. No source was found above 3$\sigma$ significance.
    
    \item  The BAT survey data, which consisted of continuous data binned in 300\,s bins. Again, no signal was detected at the source locations at $>3\sigma$ significance.

\end{itemize}

In summary, no significant ($\gtrsim 3\sigma$) gamma-ray source was found when searching in BAT raw data, BAT event data, and BAT survey data within T0 $\pm$ 1800\,s, where T0 is the onset time. More precise time slots in which the selected eFTCs were in the BAT FoV are reported in Table\,\ref{tab: multi-wavelength}.

\subsubsection*{{\it Fermi}}

We searched for GRB counterparts detected by the {\it Fermi} Gamma-ray Burst Monitor \citep[GBM,][]{Meegan2009} during the optical transients presented in this work. The GBM is composed of 12 sodium iodide (NaI) and two bismuth germanate (BGO) scintillation detectors, covering respectively the energy range 8 keV-1 MeV and 200 keV-40 MeV, with a field of view $>$ 8\,sr. We searched for counterparts in the Fermi GBM Burst Catalog \footnote{\url{https://heasarc.gsfc.nasa.gov/W3Browse/fermi/fermigbrst.html}}, that lists all triggers observed that have been classified as GRBs. For completeness, we also searched in the {\it Fermi} GBM Trigger Catalog\footnote{\url{https://heasarc.gsfc.nasa.gov/W3Browse/fermi/fermigtrig.html}}, that lists all triggers (even not classified as GRBs) observed by one or more of the 14 GBM detectors, and also in the Subthreshold Catalog\footnote{\url{https://gcn.gsfc.nasa.gov/fermi\_gbm\_subthresh\_archive.html}}. 

Table \ref{tab: multi-wavelength} reports the results of the research in the catalogues, with a time window of $+/-$ 1 day from the onset of the optical transient. No gamma-ray sources were found spatially and temporally coincident with the 10 optical transient reported in Table \ref{tab: shortlisted candidates}. However, we found three transients, DWF17c, DWF17f and DWF17g, spatially located within the 3$\sigma$ contour plot of the GBM localization of a GRB (GRB170208). This GBM event occurred on 2017-02-08 at 18:11:16.397, more than 1\,day after the optical transients, and is separated by more than 10\,degrees from the positions of the optical transients.
 Due to the time-lag between the optical and gamma-ray events and to the poor localization of the GBM detectors, we conclude that GRB170208 cannot be the counterpart to DWF17c, DWF17f or DWF17g.

\subsection{Searches for coincident fast radio bursts}
\label{sec: Parkes and Molonglo}

{\it Parkes} -- as part of the SUrvey for Pulsars and Extragalactic Radio Bursts (SUPERB) \citep{Keane2018}, we explored the data 
acquired using the 21-cm multibeam receiver \citep{Staveley-Smith1996} deployed on the Parkes radio telescope. The FWHM of each of the 13 beams is $\sim$ 14~arcmin, with an areal $\sim 2\sigma$ 13-beam coverage of $\sim 3$\,deg$^2$.

The output of each beam was processed by the Berkeley-Parkes-Swinburne
Recorder (BPSR) mode of the HI-Pulsar (HIPSR) system \citep{Keith:2010htru,
Price:2016jai, Keane2015}. The BPSR produces 8-bit, Stokes-I filterbanks with
spectral and temporal resolution of 390.625\,kHz and 64\,$\upmu$\,s respectively,
over 1182--1152\,MHz (340\,MHz bandwidth).
Searches for FRBs in the filterbanks were performed in real time using the dedicated GPU-based single pulse search software,
HEIMDALL\footnote{\href{https://sourceforge.net/projects/heimdall-astro/}{https://sourceforge.net/projects/heimdall-astro/}}.
Candidate selection criteria are listed in \S 3.3.2 of \cite{Keane2018}. No FRBs were detected in real-time with S/N $\geq 10$. A more thorough offline processing of the data with a lower S/N threshold of 6 did not yield any significant FRB detections either.\\

\noindent {\it Molonglo} -- the Molonglo radio telescope, a Mills-cross design interferometer located near Canberra, Australia, is a pulsar timing and FRB detection facility \citep{Bailes2017}. Molonglo is sensitive to right-hand circularly polarised radiation, and operates 
in the spectral range of 820-850 MHz. The relatively large ($\approx4$\,deg $\times 2.8$\,deg) primary beam of Molonglo is tiled with 352 thin synthesised `fanbeams' that overlap at their FWHM ($\approx 45$\,arcsec). 
The output of each fanbeam is an 8-bit filterbank with spectral and temporal resolution of 98 
kHz and 327 
$\upmu$s 
respectively. These filterbanks were searched for FRBs using a modified version of HEIMDALL, and burst candidates are validated via a machine learning pipeline operating in real-time \citep{Farah2018, Farah2019}.  We searched for FRBs with widths in the range 327\,$\upmu$s to 42\,ms, dispersion measures (DMs) in the range $0<$DM$<5000$\,pc\,cm$^{-3}$. 

Both Parkes and Molonglo were observing the region of sky where the selected eFTC were discovered, and the results are summarised in 
Table\,\ref{tab: multi-wavelength}. No FRBs were detected during the observations.

\subsection{Optical interleaved and long-term follow-up}
\label{subsec: skymapper}

We obtained photometry from the SkyMapper 1.35m telescope located at Siding Spring Observatory in New South Wales, Australia \citep{Keller2007}. We have established a follow-up programme which coordinates with the DWF programme to obtain interleaved nightly observations during DWF (hours-later observations once it becomes night in Australia), and late-time regularly-cadenced observations weeks after DWF to characterise long-duration transients and those that can be associated with fast transients, such as supernova shock breakouts. We obtained 100\,s exposures in \(gr\) bands centred on each DWF field, which were covered by the wide-field of view of SkyMapper (5.7 deg$^{2}$) when weather is suitable. Whenever possible, we obtained template images prior to the DWF run. We used the SkyMapper Transient Survey Pipeline \citep{Scalzo2017} to detect transients and obtain subtracted photometry. In addition to these images, we obtained photometry from the SkyMapper Supernova Survey \citep{Scalzo2017} and the Southern Sky Survey \citep{Wolf2018SM} when available for the candidates discussed in this work. These images were used to check for long term variability and are available in the \(uvgriz\) passbands up to magnitudes $i=20.27, z=19.42, u=19.34, v=19.59, g=21.57, r=21.25$ depending on the coverage of the source. Long term variability was verified from March 2016 to June 2018 for DWF17a, DWF17c, DWF17f, DWF17g and up to March 2017 for DWF17k, DWF17x and up to August 2018 for DWF17ao and DWF17ax (see Appendix \ref{appendix: SkyMapper table}).

In this search for an optical counterpart to the fast transient candidates, there were no sources at the location of the DWF transients except for two cases, DWG17f and DWF17g. DWF17f was detected in the $z$ filter on March 26th 2016 with magnitude $19.1 \pm 0.1$. DWF17g was detected more than once with SkyMapper in the \(gr\) bands on February 22nd and 27th and \(z\) band on March 26 but we did not see any significant variability in any of the observations ($g=21.0, r=20.2, z=19.1$).


\section{Rates of extragalactic fast optical transients}
\label{sec: rates}

Under the assumption that all the minute-timescale transients that we detected are of Galactic nature, we can estimate upper limits to the rates for extragalactic fast optical transients. We follow the same procedure presented in \cite{Berger2013} in order to enrich and extend the results that they obtained. The rate of extragalactic fast transients is defined as:
\begin{equation}
R_{eFT}=N/(e_{\tau} * E_A)
\end{equation}
where $N$ is the number of transient events (here, $N=3$ is defined for a non-detection, 95\% confidence), $e_{\tau}$ is the detection efficiency for the chosen timescale, 
$E_A$ is the effective areal exposure, defined as:
\begin{equation}
E_A=\textrm{FoV}_{\textrm{eff}} * (n_{\textrm{im,tot}}/n_{\textrm{im,set}})*\tau
\end{equation}

\noindent
where $\textrm{FoV}_{\textrm{eff}}$ is the effective field of view explored, $n_{\textrm{im,tot}}$ is the total number of images, $n_{\textrm{im,set}}$ is the number of images constituting the smallest set that allows exploration of the chosen timescale, and $\tau$ is the evolution timescale of the transient.  
In this work we analysed 25.76~hr of data, over 10 half-nights. We searched for transients over an effective area of $\sim 2.52$\,deg$^2$ per pointing. The efficiency of the detection pipeline is based on the test results described in \citet{Andreoni2017mary}, that returned 96.7\% completeness matching two consecutive epochs at S/N$>$10. Thus we can assume  $e_{\tau}=0.967$ for the exploration of the shortest timescale, where 2 (and only 2) detections must occur in consecutive images.  The completeness rapidly reaches $>99.9\%$ for the exploration of longer timescales thanks to the fact that a transient is selected even if `missed' in $\sim 1/3$ of the images that lie between the first and the last detection of the event on the first night it was detected (with first and last images included in the count). 

The shortest timescale that we can explore is $\tau=1.17$\,min, or 1\,min 10\,s, considering 20\,s exposures interleaved by 30\,s off-sky time.  We obtain $E_A=1.91$\,deg$^2$day and an upper limit for the areal rate of $R_{eFT} <  1.625$\,deg$^{-2}$day$^{-1}$, represented with a yellow marker in Figure\,\ref{fig:surveys}. The results for every combination of timescale and depth are plotted in Figure\,\ref{fig:surveys} and presented in the Appendix.  Results of past surveys are summarised in \citet{Berger2013} and include the Pan-STARRS1 Medium-Deep Survey \citep[PS1/MDS][]{Berger2013}, the Deep Lens Survey \citep[DLS,][]{Becker2004}, the survey of the Fornax galaxy cluster \citep{Rau2008}, ROTSE III \citep{Rykoff2005}, and MASTER \citep{Lipunov2007}.  In Figure\,\ref{fig:surveys} we consider the case in which all the selected eFTCs are Galactic in nature or spurious detections. 

\section{Discussion}
\label{sec: discussion}
This work probed the minute-timescale transient sky with deep and fast-cadenced optical observations. Such a region of the timescale-depth phase-space is accessible with only a few existing facilities and no search with the combination of area, depth, and fast cadence of DWF has ever been performed. Our searches unveiled hundreds of transient and variable events, many of which evolve in minutes.  Nine eFTC passed our specific extragalactic fast transient selection criteria, all of which are coincident with faint counterparts likely to be stellar.  

 \begin{figure}
 \centering
	\includegraphics[width=\columnwidth]{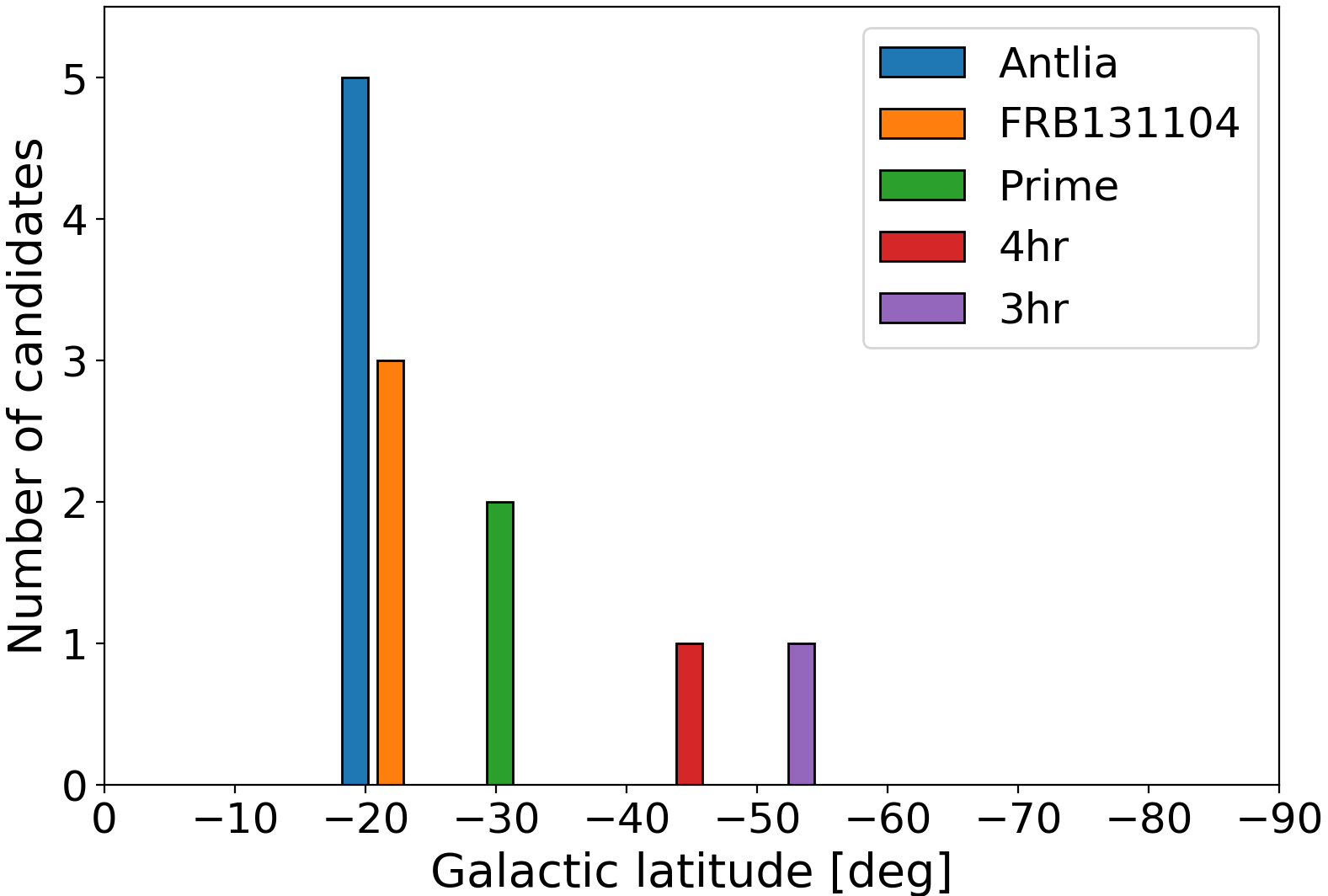}
    \caption[Number of selected candidates plotted against their Galactic latitude.]{The number of eFTCs in Table\,\ref{tab: shortlisted candidates} plotted against Galactic latitude, expressed in degrees. The bars of the histogram are 2$\degree$ wide. 
    The increase in the number of eFTCs in the Antlia field may be higher than expected as a result of the lower Galactic latitude (when extrapolating to zero latitude), or may indicate detection of extragalactic fast transients, given the proximity of the galaxy cluster.  However, we caution that the sample is too small for any definitive result. }
    \label{fig: histo Galactic latitude}
\end{figure}

Our selection criteria helped reduce contamination from Galactic sources and from transients (Galactic or extragalactic) evolving at timescales longer than $\sim 1$\,hr.  On the other hand, we are heavily biased toward discovering flare stars by imposing such strict temporal constraints.  In particular, we expect to detect `strong' flares from distant ($\gtrsim 2$\,kpc), late-type M-dwarfs difficult to detect when quiescent, as these would pass our non-stellar criteria for extragalactic events using our stellarity classifier.  Our results (see Section\,\ref{sec: results}) suggest those expectations to be correct, as most of the selected eFTCs are likely associated with late-type (usually $\geq $\,M6) stars. 

Although no deep spectra of the eFTCs exist, their peak and quiescent magnitudes and short evolution duration provide some limits on their nature when compared to known transients.
The quiescent colours and magnitudes of DWF17k, DWF17x, DWF17ao, and DWF17ax are consistent (or roughly consistent) with an M9, M9/L, M5/M6, and M6 star at $\sim 100$\,pc, $< 100$\,pc, 1000--1500\,pc, and 850--1200\,pc, respectively, and are thus thought to be flare stars.  However, their star/galaxy classifications are less robust than for other candidates.  
If the quiescent objects were instead host galaxies, their colours are inconsistent with star forming and other galaxy templates, but could be reddened elliptical or E+S0 galaxies, luminous infrared galaxies (LIRGs), or similar at $z \sim $0.5--1.5, depending on galaxy type.  The quiescent magnitudes are too bright for host galaxies at $z > 2$, placing them on the bright tip of the galaxy luminosity functions (M $\sim -23$ and brighter) before extinction correction.  

If the eFTCs are considered as fast nova-like bursts (M $\sim -8$ to $-9$), their host galaxies are constrained to M $\sim$ $-$5 to $-$9 compact dwarf galaxies or globular clusters at $\sim$ 4--11\,Mpc.  If considered as supernova shock breakouts (M $\gtrsim$ $-$20 adopted here), the hosts are constrained to $z < $0.25--0.4 and the late-time photometric limits requires any associated supernova to fainter than M $\sim -14$ to $-19$ at $\sim$100\,Mpc to $z \sim 0.3$, respectively.  The quiescent colours are inconsistent with essentially all galaxy types in this redshift range, unless heavily reddened.  Finally, if the eFTCs are optical counterparts to GRBs, the quiescent colours and magnitudes are roughly consistent with M $\sim$ $-$21 to $-$23 reddened host galaxies at $z \sim$ 0.5--1.5 and optical afterglows of M $\sim$ $-$22 to $-$23 before host or event extinction corrections.

\begin{figure}
 \centering
    \includegraphics[width=\columnwidth]{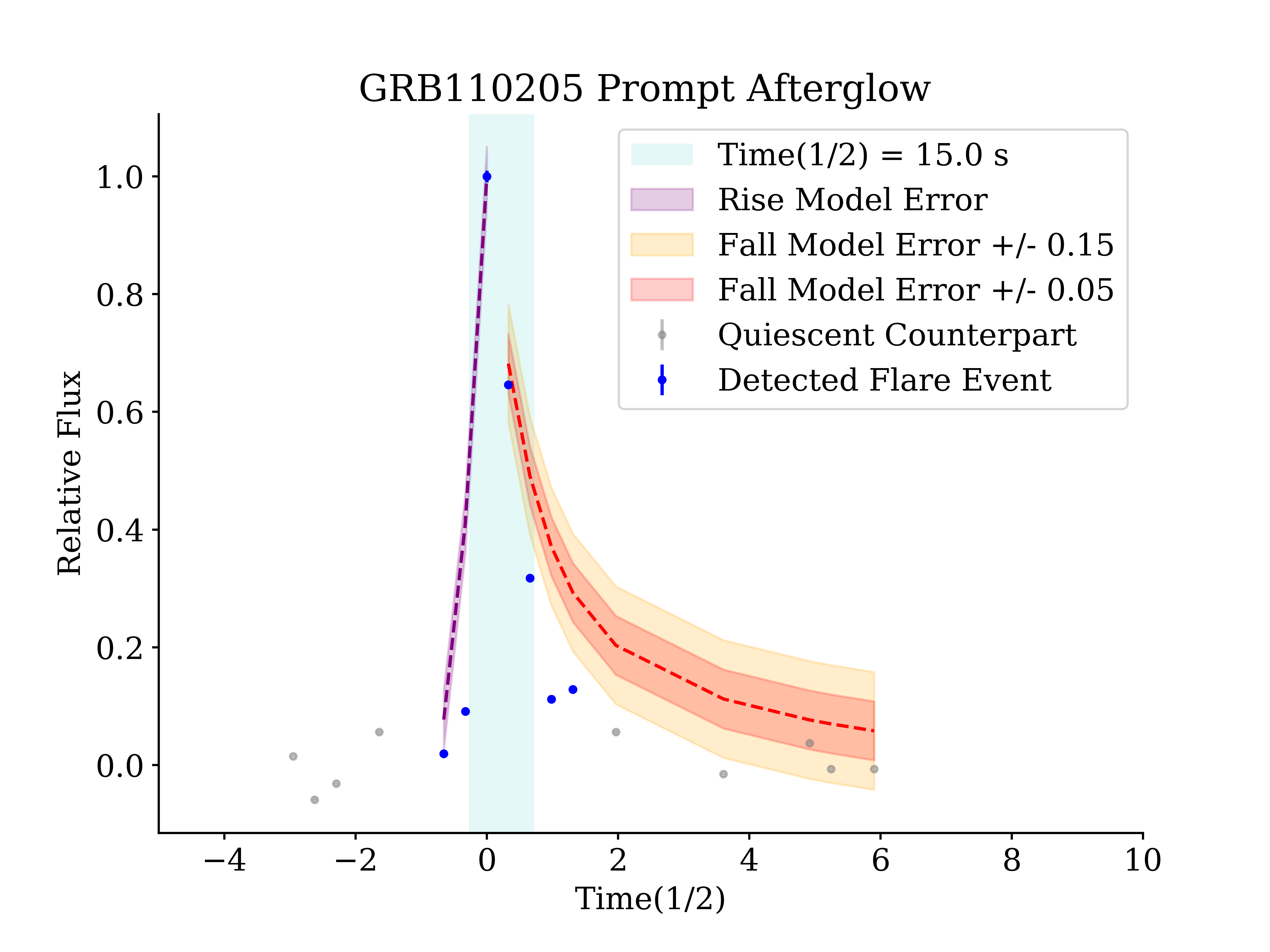}
    \caption[]{Observations of the optical flash associated with GRB110205A \citep{Cucchiara2011} are compared with the \cite{Davenport2014} flare star model described in Section\,\ref{sec: results} and Figure\,\ref{fig: flare model DWF15a DWF17a DWF17x}. }
    \label{fig: flare model GRB110205A}
\end{figure}

During searches for extragalactic fast transients, hostless candidates may be excluded because their light curves resemble flare-star light curves. To test the validity of such an argument, we compared the template flare-star model (described in Section\,\ref{sec: results}) with the light curve of the prompt optical flash that accompanied GRB\,110205A at redshift $z=2.22$ \citep{Cucchiara2011}.  The flash remained visible for less than 15 minutes, providing a good example for the type of fast transient that this work targeted. Figure\,\ref{fig: flare model GRB110205A} shows that, even without applying any re-scaling of the light curve at lower or higher redshift, the white light data points are consistent with the flare-star template. This comparison suggests that the lack of a bright host galaxy and light curve information alone cannot exclude the extragalactic nature of a fast transient candidate.

The distribution of eFTCs over Galactic latitude (Figure\,\ref{fig: histo Galactic latitude}) provides further indication that most of our eFTCs are Galactic flare stars, as they appear to be more common as the target fields approach the Galactic plane.   
A strong caveat is that Antlia is the field located at the lowest Galactic latitude among the fields considered in this analysis, but also includes the nearby Antlia galaxy cluster, which is at a comoving distance of $\sim$41\,Mpc.  Therefore our observations cannot yet exclude that minute-timescale fast optical transients are detectable with deep, fast-cadence observations both in our Galaxy and in the outskirts of nearby galaxies.    
The dependence of the number of eFTCs on Galactic latitude seems consistent with targeted flare-star population studies \citep[e.g.,][]{West2008, Kowalski2009, Hilton2010, Chang2019}. We refrain from further quantifying such dependence because of the small number of eFTCs reported in this paper, leaving it as the primary subject of a separate on-going analysis.

Finally, light curves of confirmed stellar flares show diverse behaviour, usually made more complex by a series of flares occurring in short time frames.  High time sampling can help understand the physics of flare stars and improve temporal morphology studies.
Flaring events such as the double-peak event in Figure\,\ref{fig: lc example double peak} show that poor cadence can lead to misleading interpretation of parameters such as the flare duration, without accounting for the presence of multiple peaks.  In addition, fit-to-peak estimates of peak luminosity and released energy can be overestimated using simple power-law fits.  Such overestimation can affect the study of flares individually as well as a population, while underestimating the complexity of the flaring activity.  Deep imaging and fast time sampling are necessary to compute quantities such as the duration and peak luminosity of the flares.  Future work (Webb et al., in preparation) will include detailed and complete studies of hundreds of flare stars identified during the analysis presented in this work.

\section{Summary}
\label{sec: summary}

In this paper we analysed part of the all-wavelength and all-messenger DWF programme dataset searching for extragalactic fast transients.  Hundreds to thousands of astronomical transient and variable sources were discovered in DECam optical data, 9 of which (see Table\,\ref{tab: shortlisted candidates}) passed strict selection criteria, described in Section\,\ref{subsec: selection criteria}.  Those eFTCs (extragalactic fast transient candidates) are well constrained in time, within observing windows of $\sim$1\,hr, and appear not to be coincident with stellar sources. 
Adding optical multi-band star/galaxy separation measurements to $g$-band information, we conclude that all the selected candidates are likely to have stellar progenitors.
In addition, simultaneous (and near-simultaneous) multi-wavelength observations did not identify fast radio bursts using the Parkes and Molonglo radio telescopes, or gamma-ray events associated with those eFTCs using the {\it Swift} and {\it Fermi} satellites. Those eFTC showed no significant long-term variability detectable at approximate $i=20.27, z=19.42, u=19.34, v=19.59, g=21.57, r=21.25$ magnitude limits. 

We have estimated areal rates of extragalactic fast transients at timescales ranging between 1.17 and 52.0 minutes, between $23 \lesssim g \lesssim 24.7$ survey limiting magnitudes. We assumed that all our detections are Galactic flares, which is the most likely scenario, and we place rate upper limits (Table\,\ref{tab: rates results UL}) in new regimes of the `deep and fast' region of the phase space.  

In large surveys focused on extragalactic astronomy, fast optical transients may be rejected as `contaminant' stellar flares based on their light curve. However, we showed that light curves of confirmed extragalactic fast transients (such as the prompt optical flash of GRB110205A) can mimic the behaviour of Galactic flare star light curves.  This fact suggests that more solid criteria than `hostless' and `with flare-star-like light curve' should always be adopted when searching for extragalactic transients in optical surveys. Simultaneous multi-wavelength and multi-messenger observations, along with rapid detection and prompt or long-term follow up is key to characterising the fast transient sky. Future DWF programme observations can further improve our understanding of the minute-timescale transient sky in the optical, as well as unveil the nature of the fastest bursts across several bands of the electromagnetic spectrum and using multiple messengers.

\section*{Acknowledgements}

We thank the {\it Swift} team that helped maximize the observability of DWF target fields with BAT with optimized scheduling of the satellite.  We thank Vivek Venkatraman Krishnan for helping with Molonglo operations.  We thank the SUPERB (SUrvey for Pulsars and Extragalactic Radio Bursts) team who coordinated Parkes with other DWF facilities for simultaneous observations. We thank Breakthrough Listen, who collaborated with flexible rescheduling of Parkes observing time to benefit DWF multi-facility coordination. 

\noindent
Part of this research was funded by the Australian
Research Council Centre of Excellence for Gravitational Wave Discovery
(OzGrav), CE170100004 and the Australian Research Council
Centre of Excellence for All-sky Astrophysics (CAASTRO),
CE110001020.

\noindent
Research support to IA is provided by the Australian Astronomical Observatory (AAO).  Research support to IA is also provided by the GROWTH project, funded by the National Science Foundation under Grant No 1545949. GROWTH is a collaborative project between California Institute of Technology (USA), Pomona College (USA), San Diego State University (USA), Los Alamos National Laboratory (USA), University of Maryland College Park (USA), University of Wisconsin Milwaukee (USA), Tokyo Institute of Technology (Japan), National Central University (Taiwan), Indian Institute of Astrophysics (India), Inter-University Center for Astronomy and Astrophysics (India), Weizmann Institute of Science (Israel), The Oskar Klein Centre at Stockholm University (Sweden), Humboldt University (Germany).

\noindent
JC acknowledges the Australian Research Council Future Fellowship grant FT130101219.

\noindent  FJ acknowledges funding from the European Research Council (ERC)
under the European Union’s Horizon 2020 research and innovation
programme (grant agreement No. 694745).

\noindent
This work has made use of data from the European Space Agency (ESA) mission
{\it Gaia} (\url{https://www.cosmos.esa.int/gaia}), processed by the {\it Gaia}
Data Processing and Analysis Consortium (DPAC,
\url{https://www.cosmos.esa.int/web/gaia/dpac/consortium}). Funding for the DPAC
has been provided by national institutions, in particular the institutions
participating in the {\it Gaia} Multilateral Agreement.

\noindent The national facility capability for SkyMapper has been funded through ARC LIEF grant LE130100104 from the Australian Research Council, awarded to the University of Sydney, the Australian National University, Swinburne University of Technology, the University of Queensland, the University of Western Australia, the University of Melbourne, Curtin University of Technology, Monash University and the Australian Astronomical Observatory. SkyMapper is owned and operated by The Australian National University's Research School of Astronomy and Astrophysics.

\noindent
This project used data obtained with the Dark Energy Camera (DECam),
which was constructed by the Dark Energy Survey (DES) collaboration.
Funding for the DES Projects has been provided by 
the U.S. Department of Energy, 
the U.S. National Science Foundation, 
the Ministry of Science and Education of Spain, 
the Science and Technology Facilities Council of the United Kingdom, 
the Higher Education Funding Council for England, 
the National Center for Supercomputing Applications at the University of Illinois at Urbana-Champaign, 
the Kavli Institute of Cosmological Physics at the University of Chicago, 
the Center for Cosmology and Astro-Particle Physics at the Ohio State University, 
the Mitchell Institute for Fundamental Physics and Astronomy at Texas A\&M University, 
Financiadora de Estudos e Projetos, Funda{\c c}{\~a}o Carlos Chagas Filho de Amparo {\`a} Pesquisa do Estado do Rio de Janeiro, 
Conselho Nacional de Desenvolvimento Cient{\'i}fico e Tecnol{\'o}gico and the Minist{\'e}rio da Ci{\^e}ncia, Tecnologia e Inovac{\~a}o, 
the Deutsche Forschungsgemeinschaft, 
and the Collaborating Institutions in the Dark Energy Survey. 
The Collaborating Institutions are 
Argonne National Laboratory, 
the University of California at Santa Cruz, 
the University of Cambridge, 
Centro de Investigaciones En{\'e}rgeticas, Medioambientales y Tecnol{\'o}gicas-Madrid, 
the University of Chicago, 
University College London, 
the DES-Brazil Consortium, 
the University of Edinburgh, 
the Eidgen{\"o}ssische Technische Hoch\-schule (ETH) Z{\"u}rich, 
Fermi National Accelerator Laboratory, 
the University of Illinois at Urbana-Champaign, 
the Institut de Ci{\`e}ncies de l'Espai (IEEC/CSIC), 
the Institut de F{\'i}sica d'Altes Energies, 
Lawrence Berkeley National Laboratory, 
the Ludwig-Maximilians Universit{\"a}t M{\"u}nchen and the associated Excellence Cluster Universe, 
the University of Michigan, 
{the} National Optical Astronomy Observatory, 
the University of Nottingham, 
the Ohio State University, 
the OzDES Membership Consortium
the University of Pennsylvania, 
the University of Portsmouth, 
SLAC National Accelerator Laboratory, 
Stanford University, 
the University of Sussex, 
and Texas A\&M University.
\noindent
Based on observations at Cerro Tololo Inter-American Observatory, National Optical
Astronomy Observatory 
which is operated by the Association of
Universities for Research in Astronomy (AURA) under a cooperative agreement with the
National Science Foundation.


\bibliographystyle{mnras}
\bibliography{merged} 

\appendix

\onecolumn

\section{Upper limits for areal rates}
\label{appendix: rates UL}

In this table we report upper limits for the rate of optical extragalactic fast transients.  The rates are obtained as explained in Section\,\ref{sec: rates} and are plotted in Figure\,\ref{fig:surveys}.

\begin{small}
\begin{longtable}{ p{.10\textwidth}  p{.10\textwidth}  p{.10\textwidth}  p{.10\textwidth}  p{.10\textwidth}  p{.10\textwidth}} 
\hline
\hline
 Timescale   & \multicolumn{5}{c}{Rate upper limits}\\
(minutes)   & \multicolumn{5}{c}{(events day$^{-1}$\,deg$^{-2}$)}\\
 
 \hline
    & {\bf 23.0} & {\bf 23.7} & {\bf 24.2} &{\bf 24.5} &{\bf 24.7} \\
\hline
1.17 & 1.625  & & & & \\ 
2.0 & 1.422  & & & & \\ 
2.83 & 1.294  & & & & \\ 
3.67 & 1.250  & & & & \\ 
4.5 & 1.222  & & & & \\ 
5.33 & 1.203  & & & & \\ 
6.17 & 1.189  & & & & \\ 
7.0 & 1.179  & & & & \\ 
7.83 &1.170 & 5.77  & & & \\ 
8.67 & 1.164  & & & & \\ 
9.5 & 1.158  & & & & \\ 
10.33 & 1.153  & & & & \\ 
11.17 & 1.149  & & & & \\ 
12.0 &1.146 & 5.65  & & & \\ 
12.83 & 1.143  & & & & \\ 
13.67 & 1.140  & & & & \\ 
14.5 &1.138 & & 9.81  & & \\ 
15.33 & 1.136  & & & & \\ 
16.17 &1.134 & 5.60  & & & \\ 
17.0 & 1.132  & & & & \\ 
17.83 & 1.131  & & & & \\ 
18.67 & 1.130  & & & & \\ 
19.5 & 1.128  & & & & \\ 
20.33 &1.127 & 5.56  & & & \\ 
21.17 &1.126 & & & 13.59  & \\  
22.0 &1.125 & & 9.70  & & \\ 
22.83 & 1.124  & & & & \\ 
23.67 & 1.123  & & & & \\ 
24.5 & 1.123 & 5.54  & & & \\ 
25.33 & 1.122  & & & & \\ 
26.17 & 1.121  & & & & \\ 
27.0 & 1.120  & & & & \\ 
27.83 & 1.120 & & & & 17.03  \\ 
28.67 &1.119 & 5.52  & & & \\ 
29.5 &1.119  & & 9.64  & & \\ 
30.33 & 1.118  & & & & \\ 
31.17 & 1.118  & & & & \\ 
32.0 &1.117 & & & 13.48  & \\ 
32.83 &1.117  & 5.51  & & & \\ 
33.67 & 1.116  & & & & \\ 
34.5 & 1.116  & & & & \\ 
35.33 & 1.116  & & & & \\ 
36.17 & 1.115  & & & & \\ 
37.0 &1.115  &5.50 & 9.61  & & \\ 
37.83 & 1.115  & & & & \\ 
38.67 & 1.114  & & & & \\ 
39.5 & 1.114  & & & & \\ 
40.33 & 1.114  & & & & \\ 
41.17 &1.113   & 5.49  & & & \\ 
42.0 &1.113 & & & & 16.92  \\ 
42.83 & 1.113 & & & 13.43  & \\ 
43.67 & 1.113  & & & & \\ 
44.5 &1.112 & & 9.59  & & \\ 
45.33 &1.112 & 5.49  & & & \\ 
46.17 & 1.112  & & & & \\ 
47.0 & 1.112  & & & & \\ 
47.83 & 1.112  & & & & \\ 
48.67 & 1.111  & & & & \\ 
49.5 & 1.111  & 5.48  & & & \\ 
50.33 & 1.111  & & & & \\ 
51.17 & 1.111  & & & & \\ 
52.0 & 1.111  & & & & \\ 
52.83 & 1.110  & & & & \\ 
53.67 &1.110 & 5.48  & & & \\ 
54.5 & 1.110  & & & & \\ 
55.33 & 1.110  & & & & \\ 
56.17 & 1.110 & & & & \\ 
57.0 & 1.110  & & & & \\ 
57.83 & 1.110 & & & & \\
58.67 & 1.109 & & & & \\
59.5 & 1.109 & & & & \\
\hline
\label{tab: rates results UL}
\end{longtable}
\end{small}

\section{SkyMapper photometry}
\label{appendix: SkyMapper table}
Photometry obtained by the SkyMapper telescope \citep{Keller2007} around the dates of the Deeper, Wider, Faster programmes. Photometric measurements are obtained from raw or subtracted images (see origin column) depending on template and coverage availabilityusing the SkyMapper Transient Survey pipeline \citep{Scalzo2017}. 
\begin{small}
\begin{longtable}{ p{.08\textwidth}  p{.2\textwidth}  p{.08\textwidth}  p{.08\textwidth}  p{.08\textwidth}  p{.08\textwidth}  p{.08\textwidth}} 
\hline
candidate &   date (UTC) & filter &      mag &    mag\_err &  maglim\_arr &   origin \\
\hline
   DWF17a &  2017-02-07T16:31:59 &      $g$ &        - &          - &       19.56 &    raw \\
   DWF17a &  2017-02-07T16:53:51 &      $r$ &        - &          - &       20.43 &    raw \\
   DWF17a &  2017-02-22T12:01:39 &      $g$ &        - &          - &       21.48 &    raw \\
   DWF17a &  2017-02-22T12:25:17 &      $r$ &        - &          - &       21.11 &    raw \\
   DWF17a &  2017-02-27T11:52:05 &      $g$ &        - &          - &       21.46 &    raw \\
   DWF17a &  2017-02-27T13:17:48 &      $r$ &        - &          - &       21.20 &    raw \\
   DWF17a &  2017-03-08T15:53:42 &      $g$ &        - &          - &       20.27 &    raw \\
   DWF17a &  2018-06-11T09:15:26 &      $g$ &        - &          - &       21.06 &    raw \\
   DWF17a &  2018-06-11T09:17:29 &      $g$ &        - &          - &       21.05 &    raw \\
   DWF17a &  2018-06-11T09:19:30 &      $r$ &        - &          - &       20.63 &    raw \\
   DWF17a &  2018-06-11T09:21:30 &      $r$ &        - &          - &       20.57 &    raw \\
   DWF17a &  2018-06-21T08:49:19 &      $r$ &        - &          - &       20.28 &    raw \\
   DWF17a &  2018-06-21T08:51:19 &      $g$ &        - &          - &       20.24 &    raw \\
   DWF17a &  2018-06-25T08:43:13 &      $g$ &        - &          - &       20.17 &    raw \\
   DWF17a &  2018-06-25T08:45:13 &      $r$ &        - &          - &       20.29 &    raw \\
   \hline
   DWF17c &  2017-02-07T16:31:59 &      $g$ &        - &          - &       19.28 &    raw \\
   DWF17c &  2017-02-07T16:53:51 &      $r$ &        - &          - &       20.11 &    raw \\
   DWF17c &  2017-02-22T12:01:39 &      $g$ &        - &          - &       21.42 &    raw \\
   DWF17c &  2017-02-22T12:25:17 &      $r$ &        - &          - &       20.88 &    raw \\
   DWF17c &  2017-02-27T11:52:05 &      $g$ &        - &          - &       21.42 &    raw \\
   DWF17c &  2017-02-27T13:17:48 &      $r$ &        - &          - &       21.06 &    raw \\
   DWF17c &  2017-03-08T15:53:42 &      $g$ &        - &          - &       20.17 &    raw \\
   DWF17c &  2018-06-11T09:15:26 &      $g$ &        - &          - &       21.00 &    raw \\
   DWF17c &  2018-06-11T09:17:29 &      $g$ &        - &          - &       21.00 &    raw \\
   DWF17c &  2018-06-11T09:19:30 &      $r$ &        - &          - &       20.59 &    raw \\
   DWF17c &  2018-06-11T09:21:30 &      $r$ &        - &          - &       20.48 &    raw \\
   DWF17c &  2018-06-21T08:49:19 &      $r$ &        - &          - &       20.20 &    raw \\
   DWF17c &  2018-06-21T08:51:19 &      $g$ &        - &          - &       20.13 &    raw \\
   DWF17c &  2018-06-25T08:43:13 &      $g$ &        - &          - &       20.13 &    raw \\
   DWF17c &  2018-06-25T08:45:13 &      $r$ &        - &          - &       20.21 &    raw \\
   \hline
   DWF17f &  2017-02-07T16:31:59 &      $g$ &        - &          - &       19.46 &    raw \\
   DWF17f &  2017-02-07T16:53:51 &      $r$ &        - &          - &       20.19 &    raw \\
   DWF17f &  2017-02-22T12:01:39 &      $g$ &        - &          - &       21.36 &    raw \\
   DWF17f &  2017-02-22T12:25:17 &      $r$ &        - &          - &       20.92 &    raw \\
   DWF17f &  2017-02-27T11:52:05 &      $g$ &        - &          - &       21.36 &    raw \\
   DWF17f &  2017-02-27T13:17:48 &      $r$ &        - &          - &       21.05 &    raw \\
   DWF17f &  2017-03-08T15:53:42 &      $g$ &        - &          - &       20.22 &    raw \\
   DWF17f &  2018-06-11T09:15:26 &      $g$ &        - &          - &       20.99 &    raw \\
   DWF17f &  2018-06-11T09:17:29 &      $g$ &        - &          - &       21.10 &    raw \\
   DWF17f &  2018-06-11T09:19:30 &      $r$ &        - &          - &       20.64 &    raw \\
   DWF17f &  2018-06-11T09:21:30 &      $r$ &        - &          - &       20.46 &    raw \\
   DWF17f &  2018-06-21T08:49:19 &      $r$ &        - &          - &       20.28 &    raw \\
   DWF17f &  2018-06-21T08:51:19 &      $g$ &        - &          - &       20.23 &    raw \\
   DWF17f &  2018-06-25T08:43:13 &      $g$ &        - &          - &       20.15 &    raw \\
   DWF17f &  2018-06-25T08:45:13 &      $r$ &        - &          - &       20.21 &    raw \\
   \hline
   DWF17g &  2017-02-07T16:31:59 &      $g$ &        - &          - &       19.56 &    raw \\
   DWF17g &  2017-02-07T16:53:51 &      $r$ &        - &          - &       20.43 &    raw \\
   DWF17g &  2017-02-22T12:01:39 &      $g$ &  21.02 &   0.10 &       21.48 &    raw \\
   DWF17g &  2017-02-22T12:25:17 &      $r$ &   20.18 &  0.07 &       21.11 &    raw \\
   DWF17g &  2017-02-27T11:52:05 &      $g$ &  21.24 &   0.12 &       21.46 &    raw \\
   DWF17g &  2017-02-27T13:17:48 &      $r$ &   20.21 &  0.06 &       21.20 &    raw \\
   DWF17g &  2017-03-08T15:53:42 &      $g$ &        - &          - &       20.27 &    raw \\
   DWF17g &  2018-06-11T09:15:26 &      $g$ &        - &          - &       21.06 &    raw \\
   DWF17g &  2018-06-11T09:17:29 &      $g$ &        - &          - &       21.05 &    raw \\
   DWF17g &  2018-06-11T09:19:30 &      $r$ &        - &          - &       20.63 &    raw \\
   DWF17g &  2018-06-11T09:21:30 &      $r$ &        - &          - &       20.57 &    raw \\
   DWF17g &  2018-06-21T08:49:19 &      $r$ &        - &          - &       20.28 &    raw \\
   DWF17g &  2018-06-21T08:51:19 &      $g$ &        - &          - &       20.24 &    raw \\
   DWF17g &  2018-06-25T08:43:13 &      $g$ &        - &          - &       20.17 &    raw \\
   DWF17g &  2018-06-25T08:45:13 &      $r$ &        - &          - &       20.29 &    raw \\
   \hline
   DWF17k &  2017-01-21T11:07:04 &      $g$ &        - &          - &       20.98 &    raw \\
   DWF17k &  2017-01-21T11:09:04 &      $r$ &        - &          - &       20.73 &    raw \\
   DWF17k &  2017-01-22T10:55:04 &      $g$ &        - &          - &       21.37 &    raw \\
   DWF17k &  2017-01-22T10:57:04 &      $r$ &        - &          - &       20.87 &    raw \\
   DWF17k &  2017-02-12T10:33:07 &      $v$ &        - &          - &       16.86 &    raw \\
   DWF17k &  2017-02-13T10:41:16 &      $v$ &        - &          - &       18.51 &    raw \\
   DWF17k &  2017-02-13T10:43:17 &      $v$ &        - &          - &       19.11 &    raw \\
   DWF17k &  2017-02-13T10:45:18 &      $v$ &        - &          - &       18.98 &    raw \\
   DWF17k &  2017-02-13T10:47:18 &      $v$ &        - &          - &       18.04 &    raw \\
   DWF17k &  2017-02-14T13:02:15 &      $v$ &        - &          - &       18.48 &    raw \\
   DWF17k &  2017-02-14T13:04:15 &      $v$ &        - &          - &       18.44 &    raw \\
   DWF17k &  2017-02-14T13:08:16 &      $v$ &        - &          - &       18.28 &    raw \\
   DWF17k &  2017-02-20T10:49:12 &      $v$ &        - &          - &       19.55 &    raw \\
   DWF17k &  2017-02-20T10:51:12 &      $v$ &        - &          - &       18.77 &    raw \\
   DWF17k &  2017-02-27T10:15:52 &      $g$ &        - &          - &       21.46 &    raw \\
   DWF17k &  2017-02-27T10:40:47 &      $r$ &        - &          - &       21.10 &    raw \\
   DWF17k &  2017-02-27T10:15:52 &      $g$ &        - &          - &       21.46 &    sub \\
   DWF17k &  2017-02-27T10:40:47 &      $r$ &        - &          - &       21.10 &    sub \\
   \hline
   DWF17x &  2017-01-21T11:07:04 &      $g$ &        - &          - &       20.81 &    raw \\
   DWF17x &  2017-01-21T11:09:04 &      $r$ &        - &          - &       20.71 &    raw \\
   DWF17x &  2017-01-22T10:55:04 &      $g$ &        - &          - &       21.27 &    raw \\
   DWF17x &  2017-01-22T10:57:04 &      $r$ &        - &          - &       20.72 &    raw \\
   DWF17x &  2017-02-12T10:33:07 &      $v$ &        - &          - &       16.87 &    raw \\
   DWF17x &  2017-02-13T10:41:16 &      $v$ &        - &          - &       18.48 &    raw \\
   DWF17x &  2017-02-13T10:43:17 &      $v$ &        - &          - &       19.07 &    raw \\
   DWF17x &  2017-02-13T10:45:18 &      $v$ &        - &          - &       18.99 &    raw \\
   DWF17x &  2017-02-13T10:47:18 &      $v$ &        - &          - &       17.83 &    raw \\
   DWF17x &  2017-02-14T13:02:15 &      $v$ &        - &          - &       18.42 &    raw \\
   DWF17x &  2017-02-14T13:04:15 &      $v$ &        - &          - &       18.41 &    raw \\
   DWF17x &  2017-02-14T13:08:16 &      $v$ &        - &          - &       18.17 &    raw \\
   DWF17x &  2017-02-20T10:49:12 &      $v$ &        - &          - &       19.40 &    raw \\
   DWF17x &  2017-02-20T10:51:12 &      $v$ &        - &          - &       18.48 &    raw \\
   DWF17x &  2017-02-27T10:15:52 &      $g$ &        - &          - &       21.35 &    raw \\
   DWF17x &  2017-02-27T10:40:47 &      $r$ &        - &          - &       21.00 &    raw \\
   DWF17x &  2017-02-27T10:15:52 &      $g$ &        - &          - &       21.35 &    sub \\
   DWF17x &  2017-02-27T10:40:47 &      $r$ &        - &          - &       21.00 &    sub \\
   \hline
  DWF17ao &  2017-01-22T10:51:03 &      $g$ &        - &          - &       21.57 &    raw \\
  DWF17ao &  2017-01-22T10:53:03 &      $r$ &        - &          - &       21.25 &    raw \\
  DWF17ao &  2017-02-27T10:13:52 &      $g$ &        - &          - &       21.52 &    raw \\
  DWF17ao &  2017-02-27T10:38:47 &      $r$ &        - &          - &       21.09 &    raw \\
  DWF17ao &  2017-02-27T10:13:52 &      $g$ &        - &          - &       21.52 &    sub \\
  DWF17ao &  2017-02-27T10:38:47 &      $r$ &        - &          - &       21.09 &    sub \\
  \hline
  DWF17ax &  2017-01-21T11:03:02 &      $g$ &        - &          - &       21.43 &    raw \\
  DWF17ax &  2017-01-21T11:05:03 &      $r$ &        - &          - &       20.93 &    raw \\
  DWF17ax &  2017-01-22T10:51:03 &      $g$ &        - &          - &       21.46 &    raw \\
  DWF17ax &  2017-01-22T10:53:03 &      $r$ &        - &          - &       21.23 &    raw \\
  DWF17ax &  2017-02-27T10:13:52 &      $g$ &        - &          - &       21.42 &    raw \\
  DWF17ax &  2017-02-27T10:38:47 &      $r$ &        - &          - &       21.03 &    raw \\
  DWF17ax &  2017-02-27T10:13:52 &      $g$ &        - &          - &       21.42 &    sub \\
  DWF17ax &  2017-02-27T10:38:47 &      $r$ &        - &          - &       21.03 &    sub \\
\label{tab: results SMT all}
\end{longtable}
\end{small}








\bsp	
\label{lastpage}
\end{document}